\documentclass[pdflatex,sn-nature]{sn-jnl}

\usepackage{graphicx}%
\usepackage{multirow}%
\usepackage{amsmath,amssymb,amsfonts}%
\usepackage{amsthm}%
\usepackage{mathrsfs}%
\usepackage[title]{appendix}%
\usepackage{xcolor}%
\usepackage{textcomp}%
\usepackage{manyfoot}%
\usepackage{booktabs}%
\usepackage{algorithm}%
\usepackage{algorithmicx}%
\usepackage{algpseudocode}%
\usepackage{listings}%

\usepackage[capitalise]{cleveref}%
\usepackage{natbib}%
\usepackage{subcaption}%
\usepackage{afterpage}%
\usepackage{pdflscape}%
\usepackage{makecell}%

\usepackage[acronym,section,numberedsection=autolabel]{glossaries}
\usepackage[automake]{glossaries-extra}
\makeglossaries
\newacronym{ef}{EF}{Ejection fraction}
\newacronym{cmr}{CMR}{Cardiac magnetic resonance}
\newacronym{cvd}{CVD}{Cardiovascular diseases}
\newacronym{cv}{CV}{Coefficient of Variation}
\newacronym{ed}{ED}{End-diastole}
\newacronym{edv}{EDV}{End-diastolic volume}
\newacronym{es}{ES}{End-systole}
\newacronym{esv}{ESV}{End-systolic volume}
\newacronym{lv}{LV}{Left ventricle}
\newacronym{rv}{RV}{Right ventricle}
\newacronym{myo}{MYO}{Myocardium}
\newacronym{lvef}{LVEF}{Left ventricle ejection fraction}
\newacronym{rvef}{RVEF}{Right ventricle ejection fraction}
\newacronym{sax}{SAX}{Short-axis}
\newacronym{lax}{LAX}{Long-axis}
\newacronym{2c}{2C}{Two-chamber}
\newacronym{3c}{3C}{Three-chamber}
\newacronym{4c}{4C}{Four-chamber}
\newacronym{bmi}{BMI}{Body mass index}
\newacronym{mace}{MACE}{Major adverse cardiovascular event}
\newacronym{mapse}{MAPSE}{Mitral annular plane systolic excursion}
\newacronym{gls}{GLS}{Global longitudinal shortening}
\newglossaryentry{mri}{name=MRI, description={Magnetic resonance imaging}}
\newglossaryentry{lge}{name=LGE, description={Late gadolinium enhancement}}
\newacronym{ukb}{UKB}{UK Biobank~\citep{raisi2021cardiovascular}}
\newacronym{acdc}{ACDC}{Automated cardiac diagnosis challenge~\citep{bernard2018deep}}
\newacronym{mnms}{M\&Ms}{Multi-centre, multi-vendor \& multi-disease cardiac image segmentation challenge~\citep{campello2021multi}}
\newacronym{mnms2}{M\&Ms2}{Multi-disease, multi-view \& multi-centre right ventricular segmentation in cardiac MRI~\citep{martin2023deep}}
\newacronym{kaggle}{Kaggle}{Kaggle second annual data science bowl~\citep{second-annual-data-science-bowl}}
\newacronym{rescan}{Rescan}{A multicenter, scan-rescan, human and machine learning CMR study to test generalisability and precision in imaging biomarker analysis~\citep{bhuva2019multicenter}}
\newacronym{landmark}{Landmark}{Landmark detection in cardiac MRI by using a convolutional neural network~\citep{xue2021landmark}}
\newacronym{emidec}{EMIDEC}{Evaluation of myocardial infarction from delayed-enhancement cardiac MRI~\citep{raisi2021cardiovascular}}
\newacronym{myops2020}{MyoPS2020}{Myocardial pathology segmentation challenge 2020~\citep{li2023myops}}
\newacronym{cnn}{CNN}{Convolutional neural network}
\newacronym{ci}{CI}{Confidence interval}
\newacronym{mae}{MAE}{Mean absolute error}
\newacronym{rocauc}{ROC AUC}{Area under the receiver operating characteristic curve}

\begin{document}

\title[CineMA]{A versatile foundation model for cine cardiac magnetic resonance image analysis tasks}

\author*[1,2]{\fnm{Yunguan} \sur{Fu}}\email{yunguan.fu.18@ucl.ac.uk}
\author[3,4]{\fnm{Wenjia} \sur{Bai}}
\author[1]{\fnm{Weixi} \sur{Yi}}
\author[5,6]{\fnm{Charlotte} \sur{Manisty}}
\author[5,6]{\fnm{Anish N} \sur{Bhuva}}
\author[5,6]{\fnm{Thomas A} \sur{Treibel}}
\author[5]{\fnm{James C} \sur{Moon}}
\author[1]{\fnm{Matthew J} \sur{Clarkson}}
\author[5,6]{\fnm{Rhodri Huw} \sur{Davies}}
\author[1]{\fnm{Yipeng} \sur{Hu}}

\affil*[1]{\orgname{UCL Hawkes Institute, Department of Medical Physics and Biomedical Engineering, University College London}, \country{UK}}
\affil[2]{\orgname{InstaDeep}, \country{UK}}
\affil[3]{\orgname{Department of Computing, Imperial College London}, \country{UK}}
\affil[4]{\orgname{Department of Brain Sciences, Imperial College London}, \country{UK}}
\affil[5]{\orgname{Institute of Cardiovascular Sciences, University College London}, \country{UK}}
\affil[6]{\orgname{Barts Heart Centre, Barts Health NHS Trust}, \country{UK}}

\abstract{
Cardiac magnetic resonance (CMR) is central to cardiovascular diagnosis and management, yet the extraction of key clinical measurements remains time-consuming, subjective, and limited in reproducibility.
Recent deep learning methods have enabled the automation of CMR image analysis, but each task still requires a separate model trained from scratch, demanding substantial clinical expertise to generate sufficient training data.
To facilitate the analysis of cine CMR images, here we present a versatile foundation model that can perform a range of clinically-relevant image analysis tasks, including segmentation, landmark localisation, diagnosis, and prognostication. 
A multi-view convolution–transformer masked autoencoder, named as CineMA, was trained on 15 million cine images from 74,916 subjects. 
The model was validated on multiple image analysis tasks and compared to existing models on \textgreater4,500 images from eight independent datasets with diverse population characteristics, representing the largest benchmark study for cine CMR so far.
CineMA consistently outperformed conventional convolutional neural networks (CNNs) in delineating ventricular boundaries and estimating ejection fraction, a key measure of cardiac function. The improved performance was preserved, even when the model only used half of fine-tuning data. CineMA also surpassed CNNs in disease detection and matched their performance in long-axis function measurement. 
Interestingly, we found that CineMA can also detect cardiac changes in systemic diseases, such as diabetes, hypertension and cancer, and can also predict mortality. Finally, we assessed model fairness and demonstrated consistent model performance across demographic subgroups. These findings highlight CineMA’s accuracy, learning efficiency, adaptability, and fairness, underscoring its potential as a foundation model for automated cardiac image analysis to support clinical workflow and cardiovascular research. All training and inference code and models are made publicly available at \url{https://github.com/mathpluscode/CineMA}.
}


\maketitle

\section{Introduction}
\label{sec:introduction}

Despite major advances over the last few decades, cardiovascular diseases remain the leading cause of mortality worldwide, accounting for over 19.8 million deaths annually~\citep{mensah2023heart}. Many clinical decisions in disease diagnosis are driven by imaging assessment of cardiac structure and function. Due to its excellent soft tissue contrast, non-invasiveness, and non-ionising nature, cardiovascular magnetic resonance (\gls{cmr}) is now a standard part of many clinical pathways as reflected in many international guidelines ~\citep{rajiah2023cardiac,von2023cardiovascular,10.1093/eurheartj/ehaf192}. Cine imaging, a fundamental component of cardiac MRI, enables quantification of cardiac size (e.g. cavity volume), myocardial mass, global systolic function (e.g. left ventricle ejection fraction [\gls{lvef}]), long-axis function (e.g. mitral annular plane systolic excursion [\gls{mapse}], and global longitudinal shortening [\gls{gls}]). CMR is established as the gold standard for estimating these measures and providing both diagnostic and prognostic values \citep{kramer2020standardized}. However, interpreting CMR images and deriving biomarkers for diagnosis, monitoring, and prognostication remain labour-intensive tasks that require tremendous efforts by experienced clinicians. In addition, the annotation process is inherently subjective, contributing to substantial intra- and inter-observer variability~\citep{bhuva2019multicenter}, which can affect clinical decision-making and compromise consistency in patient care.
With the advancement of deep learning, automated image interpretation has become feasible, helping to alleviate clinician workload and improve clinical workflow efficiency ~\citep{bernard2018deep,campello2021multi,martin2023deep,second-annual-data-science-bowl,bhuva2019multicenter,xue2021landmark,lalande2020emidec,li2023myops}.

One particular advancement in deep learning is the foundation models~\citep{bommasani2021opportunities}, models trained on large datasets and adaptable to diverse downstream tasks. Such models hold great potential for CMR imaging analyses, which involve diverse tasks from segmentation, landmark localisation, to diagnosis and prognostication. MedSAM~\citep{ma2024segment} is one example of the foundation model, developed for general 2D medical image segmentation. However, it is not fully automated, requiring manually provided bounding box prompts to perform segmentation. Also, it is not able to perform various CMR analysis tasks. 
Sun et al.~\citep{sun2025foundation} developed a foundation model for correcting motion, enhancing resolutions, and denoising brain MR images. The enhanced images yielded improved performance in downstream tasks, including tissue segmentation, registration, and diagnosis. Several foundation models tailored specifically to CMR have recently been proposed~\citep{shad2023generalizable,jacob2024towards,zhang2025towards}. Shad et al.~\cite{shad2023generalizable} used 14,073 cardiac MRI scans to align vision and text encoders, and then fine-tuned the vision encoder for LVEF prediction and disease detection on the \gls{acdc} dataset~\citep{bernard2018deep}. Jacob et al.~\cite{jacob2024towards} introduced a 2D model trained on 27,524 CMR studies using the DINO pre-training framework~\citep{caron2021emerging}, and evaluated it for disease detection, ventricle segmentation, and landmark localisation tasks across multiple datasets, including ACDC, \gls{kaggle}~\citep{second-annual-data-science-bowl}, and \gls{emidec}~\citep{lalande2020emidec}. More recently, Zhang et al.~\cite{zhang2025towards} proposed a multi-modal foundation model by training a multi-view masked autoencoder~\citep{he2022masked} on 42,000 CMR studies, and aligning image embeddings with tabular clinical features using contrastive learning. Their model was evaluated on UK Biobank (\gls{ukb}) data across multiple tasks, including ejection fraction (\gls{ef}) estimation, disease classification, and segmentation. These studies provide the first evidence of the benefits from models pre-trained with growing cine CMR data. However, their evaluations were limited in several aspects, such as the absence of convolutional neural network baselines~\citep{caron2021emerging,zhang2025towards}, the omission of clinically relevant metrics, and the lack of validation on external datasets~\citep{zhang2025towards}. While the reported results are promising, the proposed models were not consistently comparable to strong convolutional baselines.

Here we present CineMA (Cine CMR Masked Autoencoder), a cine CMR foundation model using the self-supervised masked autoencoder framework~\citep{he2022masked}, trained on 74,916 cine CMR studies comprising over 15 million images from the UKB. We hypothesised that large-scale pre-training would yield transferable representations for downstream tasks. Fine-tuned variants of CineMA were benchmarked against convolutional neural networks (\gls{cnn}) across a range of clinically relevant tasks. These tasks were selected to represent key components of cardiac image interpretation in routine practice, including segmentation, landmark localisation, diagnosis, and prognostication. CineMA consistently outperformed CNNs, especially in low-data settings and under population shifts. It achieved higher segmentation accuracy, more precise EF estimation, and improved specificity in disease diagnosis, suggesting potential for triage and stratification. CineMA matched CNNs in long-axis function estimation, demonstrating adaptability. Evaluated across eight datasets with over 4,500 images~(\Cref{tab:data_labels}), it is the first foundation model to achieve superior performance to CNNs and previous methods across such a broad range of cine CMR tasks~\citep{shad2023generalizable,jacob2024towards,bhuva2019multicenter,xue2021landmark}. Beyond benchmarking, CineMA-derived metrics were associated with non-cardiac conditions (diabetes, hypertension, cancer), long-term survival, and showed small disparities across demographic subgroups. These analyses highlight the broader applicability and trustworthiness of CineMA in real-world populations and support the use of the proposed foundation models over task-specific training for future cardiac imaging applications. All code for pre-training, fine-tuning, and inference, along with pre-trained and fine-tuned models, is publicly available at \url{https://github.com/mathpluscode/CineMA}, enabling reproducibility and direct application of the foundation model.

\clearpage
\section{Results}
\subsection{Model architectures and experimental setting}
Following Zhou et al.~\cite{zhou2023foundation}, we adapted the MAE framework~\citep{he2022masked} for model pre-training by reconstructing images from masked inputs. To support multiple views of cine CMR including long-axis (\gls{lax}) and short-axis (\gls{sax}) views with a single unified backbone model, we adopted the MultiMAE architecture~\citep{bachmann2022multimae}, in which each view is encoded and decoded independently, before and after a shared Transformer encoder. Additionally, we incorporated convolutional layers into the encoding and decoding stages to reduce memory consumption and improve training efficiency ~\cite{gao2022convmae}. The proposed multi-view convolution–transformer masked autoencoder model is named CineMA, short for cine CMR masked autoencoder (\Cref{fig:summary}).
We collected 74,916 cine CMR studies from the UK Biobank (up to November 6, 2023)~\citep{raisi2021cardiovascular}, comprising more than 15 million images in total. CineMA was pre-trained on this dataset, then fine-tuned on eight datasets for a range of downstream segmentation and diagnostic tasks~(\Cref{tab:data_labels}). During fine-tuning, encoder branches corresponding to unavailable views and decoder layers were removed. Task-specific heads were added depending on the downstream labels: UNetR-style architectures~\citep{hatamizadeh2022unetr} were employed for segmentation masks, and linear prediction layers were used for tabular labels such as disease diagnosis and clinical measurements~(\Cref{fig:summary}). We compared the performance of fine-tuned CineMA (denoted CineMA\textsuperscript{FineTune}) against two baselines: (1) a model with the same architecture as CineMA\textsuperscript{FineTune} but trained from random initialisation (denoted CineMA\textsuperscript{RandInit}), and (2) a convolutional neural network trained from random initialisation, UNet for segmentation labels (denoted UNet\textsuperscript{RandInit}) and ResNet-50~\citep{targ2016resnet} (denoted ResNet\textsuperscript{RandInit}) for tabular labels. For each downstream task experiment, we trained the model using three different random seeds, and the evaluation metrics were averaged across the three runs, including Dice score for segmentation, area under the receiver operating characteristic curve (\gls{rocauc}) for disease classification, and mean absolute error (\gls{mae}) for diagnostic regression. Bootstrapping on the test set was applied to perform two-sided $t$-tests to assess statistical significance. Significance levels are denoted as follows: *** for $p < 0.001$, ** for $p < 0.01$, * for $p < 0.05$, and \textit{ns} for non-significant differences.

\begin{figure}[!htbp]
    \centering
    \includegraphics[height=\textheight]{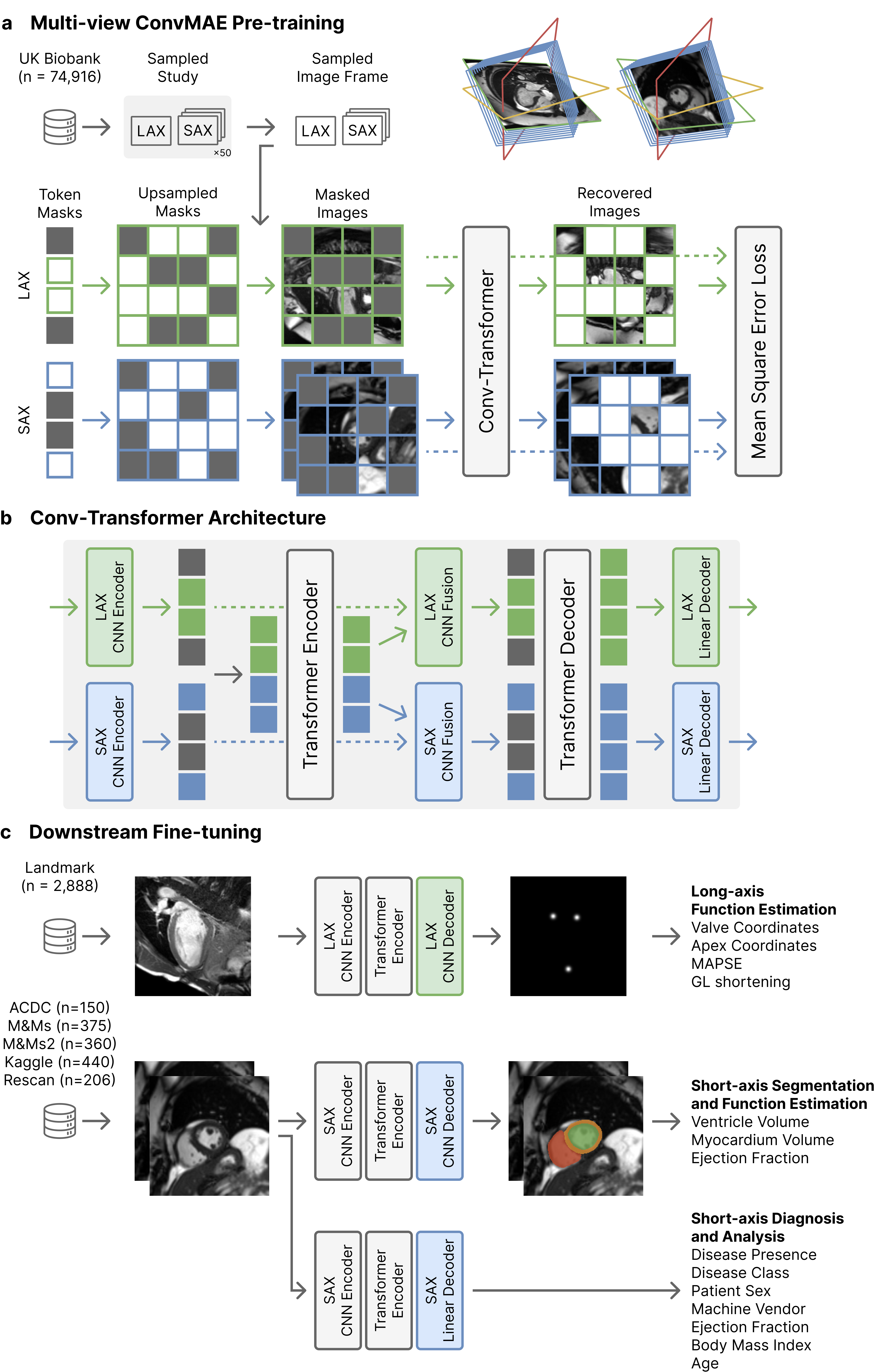}
\end{figure}
\begin{minipage}{\textwidth}
\captionof{figure}{CineMA is a multi-view convolution–transformer masked auto-encoder. \textbf{a}, CineMA was pre-trained on 74,916 cine cardiac MR studies. One image frame is sampled from a randomly selected study. Views are randomly masked and reconstructed using Conv-Transformer.  Mean squared error loss is computed per view and then averaged. All three LAX views (two-chamber, three-chamber, and four-chamber) were used, but only the two-chamber view is illustrated for simplicity. \textbf{b}, Conv-Transformer architecture. Each masked view is independently encoded by convolutional neural networks (CNNs). Visible patch tokens are concatenated and passed to a Transformer encoder. After multi-scale fusion, visible and masked tokens are decoded by a Transformer decoder to reconstruct each view. \textbf{c}, CineMA was fine-tuned separately on multiple datasets to perform diverse clinically-relevant tasks. During fine-tuning, pre-trained decoders were discarded and replaced by task-specific heads. SAX, short-axis. LAX, long-axis. MAPSE, mitral annular plane systolic excursion. GLS, global longitudinal shortening.}
\label{fig:summary}
\end{minipage}

\subsection{Fine-tuning CineMA improves ventricle delineation and EF assessment over CNNs trained from scratch}
\begin{figure}[!htbp]
\centering
\includegraphics[width=0.8891428571\textwidth]{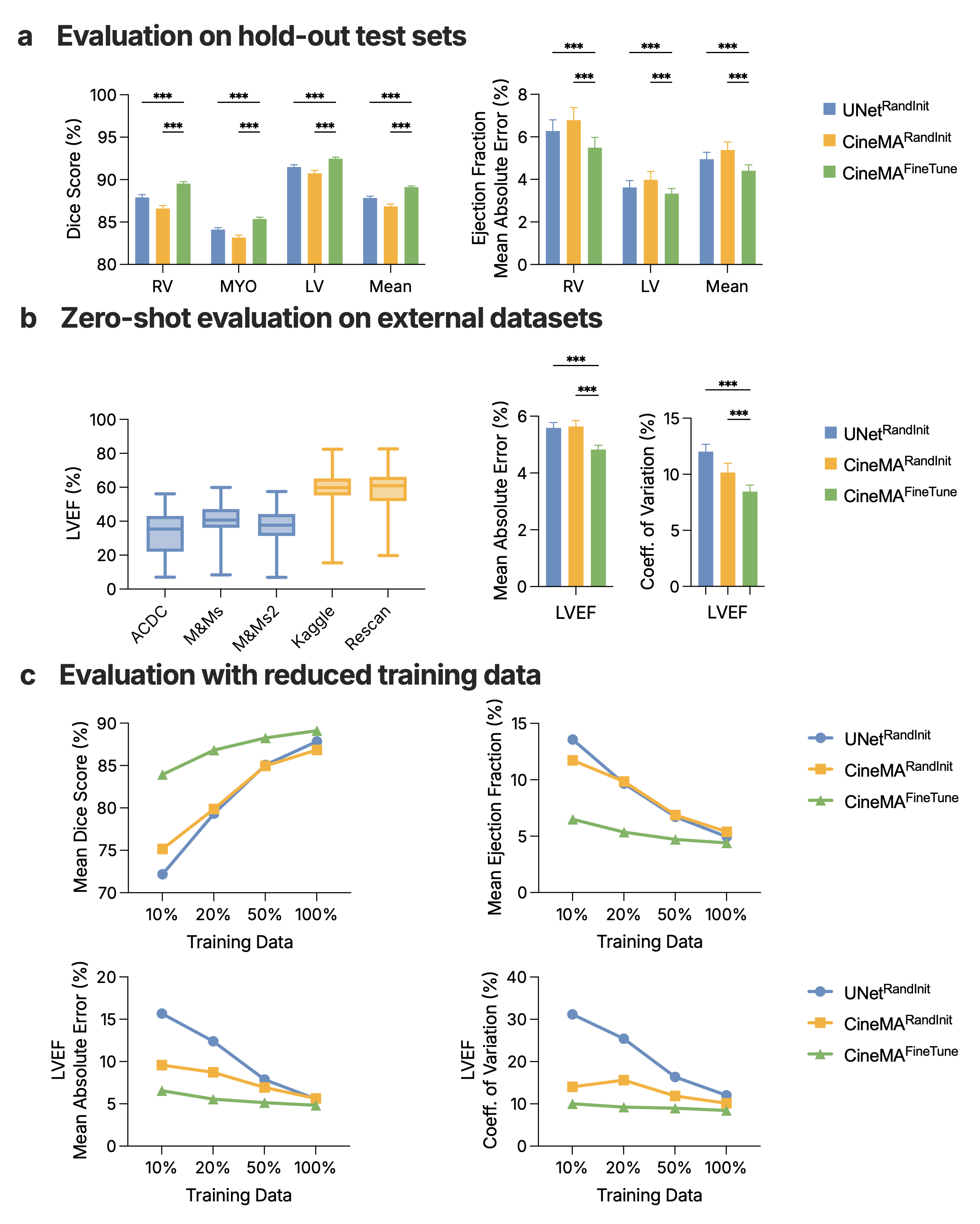}
\caption{Segmentation model evaluation on combined hold-out test sets from ACDC, M\&Ms, and M\&Ms2 datasets, and on external Kaggle and Rescan datasets. \textbf{a}, Fine-tuned CineMA (green) demonstrated significantly higher Dice scores for segmentation and lower errors in ejection fraction prediction. \textbf{b}, Kaggle and Rescan datasets exhibit different left ventricular ejection fraction distributions compared to the training datasets, indicating a population shift. Fine-tuned CineMA achieved lower errors and higher coefficients of variation. \textbf{c}, Fine-tuned CineMA consistently outperformed baselines when using a fraction of training data (10\%, 20\%, 50\%), demonstrating superior sample efficiency. RV, right ventricle. MYO, myocardium. LV, left ventricle. RV, right ventricle. EF, ejection fraction. Coeff., coefficient.}
\label{fig:cine_segmentation_combined}
\end{figure}
\begin{figure}[!htbp]
\centering
\includegraphics[width=\textwidth]{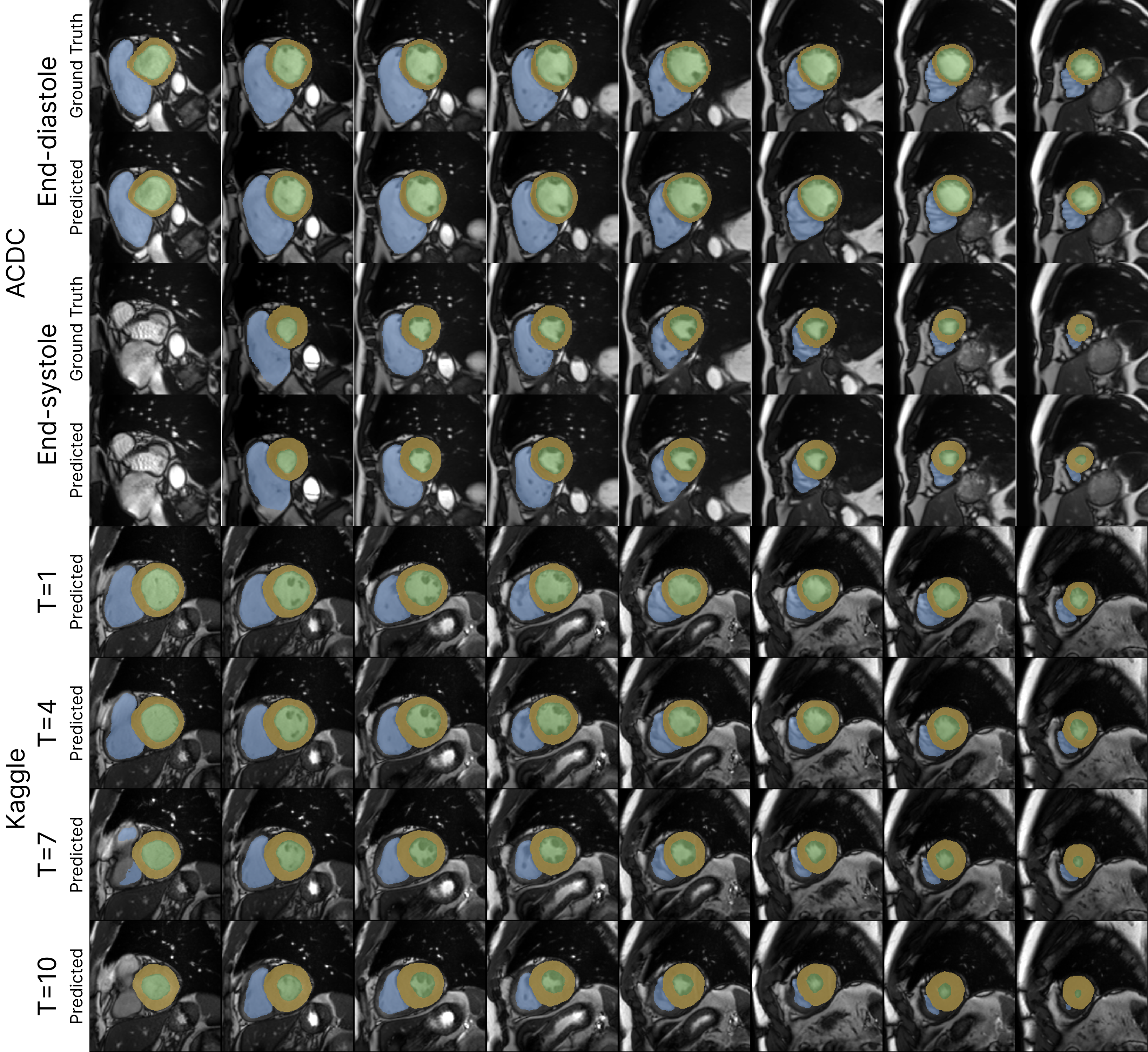}
\caption{Example segmentations on ACDC and Kaggle dataset test splits. The segmentation model was a fine-tuned version of CineMA using ACDC dataset. The model was not trained on Kaggle dataset. Each column corresponds to a short-axis slice.}
\label{fig:example_segmentation}
\end{figure}

We first evaluated CineMA by fine-tuning it for ventricular and myocardial segmentation on three independent short-axis cine CMR datasets (ACDC~\citep{bernard2018deep}, \gls{mnms}~\citep{campello2021multi}, and \gls{mnms2}~\citep{martin2023deep}). Across these datasets, CineMA\textsuperscript{FineTune} achieved the highest mean Dice score of 89.11\%, with improvements most pronounced for the right ventricle. On the ACDC benchmark, Dice scores reached over 91.06\% for the right ventricle (\gls{rv}), 87.93\% for the myocardium (\gls{myo}), and 93.56\% for the left ventricle (\gls{lv}), outperforming the previously reported scores of 90.7\%, 87.9\%, and 93.3\%, respectively, from Jacob et al.~\cite{jacob2024towards}~(\Cref{tab:supp_cine_segmentation}). CineMA\textsuperscript{FineTune} also produced the lowest Hausdorff distances, reflecting superior boundary delineation and anatomical consistency (\Cref{fig:example_segmentation}).
The gains in segmentation accuracy translated into more precise ventricular volume estimation, with mean absolute errors of 7.0 ml for the left ventricle and 10.4 ml for the right ventricle. When comparing end-diastolic and end-systolic volumes, CineMA\textsuperscript{FineTune} produced accurate ejection fraction estimates, with errors of 3.33\% for LVEF and 5.49\% for RVEF. These results highlight CineMA’s ability to deliver reliable ejection fraction estimates through accurate segmentation, supporting its use for automated and reproducible cardiac function assessment in clinical practice.

\subsection{LVEF prediction generalises across different populations without additional training}

The fine-tuned CineMA was further evaluated in a zero-shot setting by segmenting the left ventricle in the Kaggle~\citep{second-annual-data-science-bowl} and \gls{rescan}~\citep{bhuva2019multicenter} datasets, without additional training. LVEF was then derived from maximum and minimum LV volumes over the cardiac cycle. These datasets, which were not used during pre-training or fine-tuning, represented a marked population shift, with average LVEF more than 20\% higher than in the training data population (\Cref{fig:cine_segmentation_combined}). CineMA\textsuperscript{FineTune} achieved the lowest mean absolute error of 4.83\% for LVEF, with performance across fine-tuned variants ranging from 3.77\% to 5.21\%, substantially outperforming baseline models and also the 6.88\% error reported in previous work~\citep{shad2023generalizable}~(\Cref{tab:supp_cine_segmentation_ef_ood}).

The Rescan dataset consists of paired images representing two independent scans of the same subject acquired within a short interval (within one week in 96\% of cases). LVEF can therefore be assumed to remain consistent across repeated acquisitions, allowing assessment of the model’s consistency using the coefficient of variation (\gls{cv}). CineMA\textsuperscript{FineTune} achieved a CV of 8.45\% across paired scans, lower than CineMA\textsuperscript{RandInit} (10.16\%) and UNet\textsuperscript{RandInit} (12.02\%). CineMA\textsuperscript{FineTune} also outperformed the previously reported CV of 8.8\% by Bhuva et al.~\citep{bhuva2019multicenter}, despite not being trained on the Rescan dataset. These reductions in both error and variability demonstrate CineMA’s ability to perform accurate and consistent EF estimation on previously unseen data distributions, despite substantial population shift. This supports the potential utility of foundation models in real-world clinical workflows, particularly for longitudinal patient monitoring.

\subsection{EF accuracy is preserved with half the labels and generalises across populations}
The prior knowledge gained during pre-training was further assessed through sample efficiency, that is, the amount of labelled data required to achieve a given level of performance. By varying the proportion of training data, we found that CineMA\textsuperscript{FineTune} consistently outperformed both CineMA\textsuperscript{RandInit} and UNet\textsuperscript{RandInit} across all metrics and datasets, including population-shifted external cohorts~(\Cref{fig:cine_segmentation_combined}). With only half of the available training data, CineMA\textsuperscript{FineTune} achieved mean errors of 3.72\% for LVEF and 5.72\% for RVEF, matching or exceeding the accuracy of a fully trained UNet. On the Kaggle and Rescan datasets, strong performance was maintained with as little as 20\% of the training data, yielding an LVEF MAE of 5.56\% and a coefficient of variation of 9.24\%, both superior to the fully trained UNet. These results highlight CineMA’s capacity to deliver accurate segmentation and EF estimation even in low-data settings, reducing manual annotation required for fine-tuning while preserving generalisation across populations.

\subsection{Diagnostic performance extends to disease diagnosis and long-axis function evaluation}
\begin{figure}[!htbp]
\centering
\includegraphics[width=\textwidth]{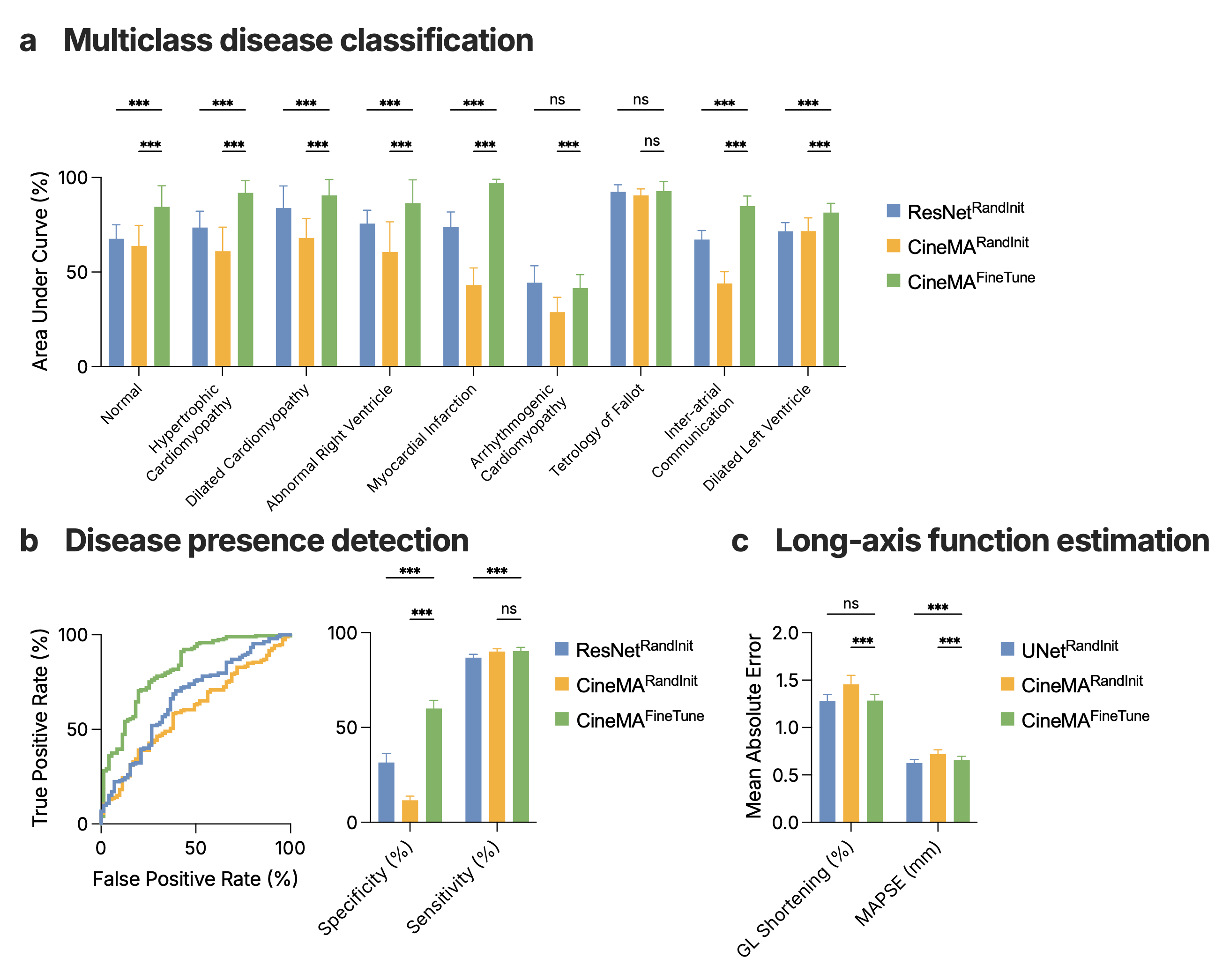}
\caption{Diagnostic performance evaluation on diseases classification, detection and long-axis function estimation. \textbf{a}, Fine-tuned CineMA significantly outperformed ResNet\textsuperscript{RandInit} across almost all cardiovascular disease and normal conditions. \textbf{b}, Fine-tuned CineMA consistently achieved higher sensitivity, leading to higher area under the curve (AUC) values, and also demonstrated higher specificity, indicating potential to reduce over-diagnosis while maintaining a low false-negative rate. \textbf{c}, Fine-tuned CineMA achieved performance comparable to UNet\textsuperscript{RandInit} for long-axis function estimation. MAPSE, mitral annular plane systolic excursion. GLS, global longitudinal shortening.}
\label{fig:diagnosis}
\end{figure}
Beyond ventricle segmentation and ejection fraction estimation, we evaluated CineMA’s diagnostic performance across multiple tasks, including cardiovascular disease diagnosis and long-axis function assessment. Disease classification was performed separately on the ACDC, M\&Ms, and M\&Ms2 datasets, each comprising distinct diagnostic labels. CineMA\textsuperscript{FineTune} consistently outperformed baseline models across all three datasets (\Cref{tab:supp_cvd_classification}), with gains observed across nearly all disease classes, including the commonly observed hypertrophic cardiomyopathy (\Cref{fig:diagnosis}). For general disease detection, cardiac disease types were merged into a single ``disease present'' category. CineMA\textsuperscript{FineTune} achieved higher specificity (60.06\% vs. 31.55\%) and sensitivity (90.20\% vs. 86.84\%) than ResNet\textsuperscript{RandInit}(\Cref{fig:diagnosis}, \Cref{tab:supp_cvd_classification}). This improvement in specificity suggests CineMA\textsuperscript{FineTune} could reduce over-diagnosis while maintaining a low false-negative rate, supporting its potential role in automated screening and triage.
CineMA\textsuperscript{FineTune} was also fine-tuned for image-wise classification and regression tasks, including MRI vendor, patient sex, EF, BMI, and age~(\Cref{tab:supp_vendor_sex_classification}, \Cref{tab:supp_regression}). Across all tasks, CineMA\textsuperscript{FineTune} demonstrated that self-supervised pre-training yields transferable representations for both clinical and acquisition-related characteristics.

Long-axis function was assessed on two-chamber (\gls{2c}) and four-chamber (\gls{4c}) views using the \gls{landmark} dataset, by localising keypoints at the mitral valve and apex that define the mitral valve plane and ventricular length ~\citep{xue2021landmark}. Heatmap regression was used to predict dense probability maps of landmark locations. CineMA\textsuperscript{FineTune} achieved mean Euclidean distance errors of 0.90~mm in 2C view and 1.10~mm in 4C views, improving on previously reported errors~\citep{xue2021landmark} by at least 1mm~(\Cref{tab:supp_landmark}). For MAPSE estimation, CineMA\textsuperscript{FineTune} achieved an MAE of 0.66~mm, slightly higher than the UNet\textsuperscript{RandInit} baseline (0.63~mm), though the difference is clinically negligible. GLS estimation yielded identical MAEs of 1.28\% for both CineMA\textsuperscript{FineTune} and UNet\textsuperscript{RandInit}, indicating comparable accuracy in these functional assessment tasks.

\subsection{CineMA detects changes in cardiac function in diabetes, hypertension, and cancer}
\begin{figure}[!htbp]
\centering
\includegraphics[width=0.8\textwidth]{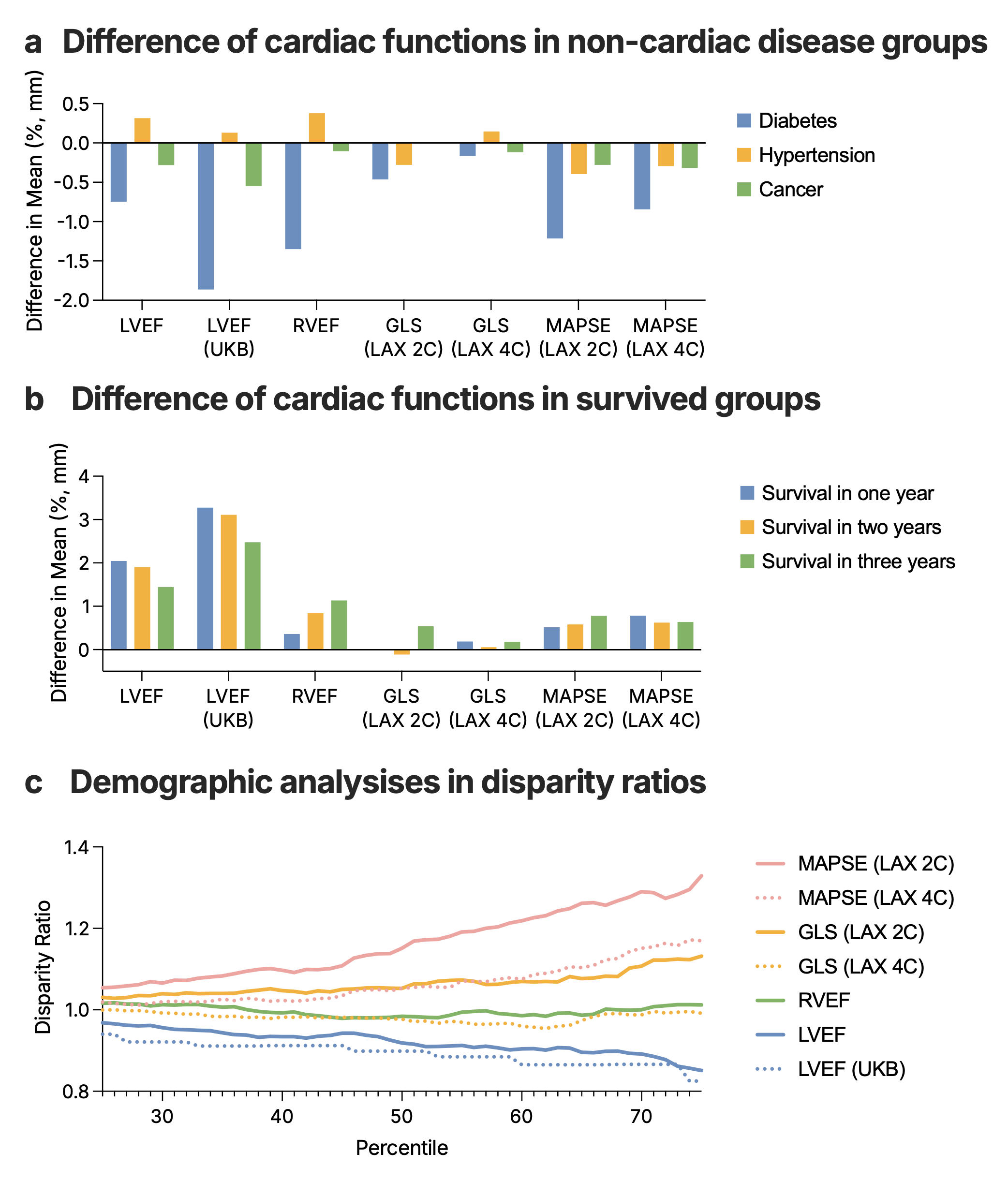}
\caption{Association and demographic analyses with predicted cardiac functions. \textbf{a}, Diabetes and cancer were associated with reduced cardiac function, while hypertension was associated with higher ejection fraction. \textbf{b}, Surviving individuals exhibited better cardiac function across metrics. \textbf{c}, CineMA-predicted LVEF showed disparity ratios consistently closer to 1.0 compared to the LVEF derived from UKB Heart MRI labels (inlineVF), suggesting reduced demographic bias.}
\label{fig:stats_analysis}
\end{figure}
As CineMA\textsuperscript{FineTune} models achieved high accuracy and consistency across cardiac tasks, we next examined associations between predicted cardiac function and systemic diseases in the UKB cohort (\Cref{fig:stats_analysis}). After adjusting for age, sex, and BMI, individuals with diabetes showed consistently lower ventricular systolic performance compared to those without diabetes (\Cref{tab:diabetes}). The largest reduction was seen in left and right ventricular ejection fraction ($\beta=-0.41\%$, $p=0.002$; $\beta=-0.69\%$, $p < 0.001$) and in long-axis functional measures, particularly MAPSE from the 2-chamber ($\beta=-1.01$~mm, $p < 0.001$) and 4-chamber ($\beta=-0.57$~mm, $p < 0.001$) views. The effect detected with CineMA-predicted LVEF was smaller than the $1.33\%$ reduction observed using LVEF values automatically derived from inlineVF (available for 53.6\% of the cohort). Overall, CineMA’s predictions indicate modest diabetes-related myocardial dysfunction.
In contrast, hypertension was associated with higher ejection fraction (e.g., $\beta=0.59\%$ for LVEF, $p < 0.001$; $\beta=0.90\%$ for RVEF, $p < 0.001$), with weaker or non-significant associations for long-axis function (\Cref{tab:blood_pressure}), suggesting a distinct physiological pattern compared with diabetes.
Cancer, whether diagnosed before or after imaging, was similarly associated with reduced ventricular systolic performance across multiple parameters (\Cref{tab:cancer}). The most pronounced reduction was observed in left and right ventricular ejection fraction ($\beta=-0.39\%$, $p < 0.001$; $\beta=-0.28\%$, $p < 0.001$). Reductions were also seen in MAPSE from 2-chamber and 4-chamber LAX view ($\beta=-0.12$~mm, $p=0.001$; $\beta=-0.13$~mm, $p < 0.001$), indicating subtle impairments in myocardial contractility in individuals with cancer.

\subsection{Prognostic association between cardiac function and survival}
We next assessed the prognostic value of CineMA-derived cardiac functional metrics for predicting 1-, 2-, and 5-year survival (\Cref{fig:stats_analysis}). In Cox proportional hazards models adjusted for age, sex, and BMI, CineMA-predicted LVEF was significantly associated with survival at all time points, with effect sizes and significance levels comparable to UK Biobank provided LVEF (Table~\ref{tab:survival}). Each 1\% reduction in LVEF corresponded to an estimated 3–5\% increase in mortality risk. While RVEF and MAPSE were not predictive of short-term survival, both became significant predictors at 5 years (both $p < 0.05$), suggesting delayed but clinically relevant prognostic value. These findings highlight CineMA’s potential to provide accurate cardiac function estimates that carry long-term prognostic information.

\subsection{Demographic disparity analysis}

UK Biobank data predominantly consists of White subjects (90.86\%), making it important to evaluate whether models exhibit ethnic bias. Because no ground-truth labels for cardiac functions are available, we assumed that a fair model would predict a positive cardiac function outcome, such as LVEF exceeding 40\%, with equal probability across ethnic groups~\citep{seyyed2020chexclusion}. We computed the disparity ratio of positive outcome rates between White and non-White subjects. Because cardiac function metrics span different value ranges, we defined thresholds ranging from the 25th to 75th percentiles of the predicted values across all subjects. As shown in \Cref{fig:stats_analysis}, CineMA-derived LVEF exhibited disparity ratios ranging from 0.85 to 0.97, suggesting relatively low disparity across ethnic groups. Notably, CineMA-derived LVEF yielded disparity ratios that were always closer to 1.0 than those based on the UKB-provided LVEF labels.
CineMA-predicted RVEF showed smaller disparities (0.98–1.02) still, while among long-axis function metrics, GLS had lower disparities (0.95–1.13) and slightly higher (1.01–1.33) for MAPSE. 
To our knowledge, this is the first systematic evaluation of demographic bias in cardiac function prediction using foundation models, supporting CineMA’s potential for fair and generalisable deployment in diverse populations.

\section{Discussion}
\label{sec:discussion}
In this study, we introduced CineMA, the first cine CMR foundation model developed for a broad range of clinically relevant tasks. CineMA is a multi-view convolution–transformer masked autoencoder that was pre-trained on over 15 million CMR images and evaluated on eight independent datasets. CineMA demonstrated superior segmentation accuracy and precise and consistent estimates of ejection fraction, particularly in low-data scenarios and in external datasets with population shifts. CineMA also performed well in disease detection and diagnosis, achieving substantially higher specificity and increased sensitivity compared to the baseline methods, potentially reducing over-diagnosis. Through this comprehensive benchmark, we demonstrated the value of the proposed cine CMR foundation models in reducing training data requirements and improving performance, consistency, and generalisability. These findings highlight the advantages of fine-tuning foundation models over task-specific training and their potential to enable cardiac imaging applications in settings previously constrained by limited data or suboptimal model performance. To assess CineMA’s clinical relevance beyond task benchmarking, we analysed associations between predicted cardiac function metrics and systemic disease as well as survival outcomes. We identified modest but significant reductions in systolic and long-axis function among patients with diabetes or cancer. Lower CineMA-predicted LVEF was also consistently associated with increased mortality risk, demonstrating its prognostic value. Finally, using UK Biobank data, we analysed the disparities in model errors across demographic subgroups, supporting CineMA’s fairness and robustness. Together, these results highlight CineMA’s potential not only for technical performance but also for large-scale epidemiological research, risk stratification, and equitable clinical deployment.

Multiple prior efforts have explored cardiac foundation models by fine-tuning existing models~\cite{shad2023generalizable} or pre-training new ones~\cite{jacob2024towards,zhang2025towards}. These models have evolved from 2D to multi-view architectures, with dataset sizes growing to as many as 42,000 CMR studies. CineMA advances this direction by incorporating convolutional layers to reduce memory consumption and improve performance, and by scaling pre-training to over 74,000 CMR studies comprising more than 15 million images. While these works have demonstrated preliminary evidence of the benefits of foundation models, their evaluations were often fragmented, performed under varying conditions and lacked standardisation in baselines, metrics, or datasets. In particular, some omitted clinically relevant metrics, lacked strong convolutional neural network baselines, or were not validated on external datasets. Additionally, prior studies did not investigate associations with systemic diseases or clinical outcomes, nor did they assess fairness across patient subgroups. To address these gaps, our study consolidated a broad set of clinically relevant tasks and extended the benchmark to include more external datasets, assess additional metrics such as prediction consistency across repeated scans, and systematically examine sample efficiency by varying training data sizes. To our knowledge, this is the first cardiac foundation model evaluated for associations with diabetes, cancer, survival outcomes, and demographic fairness. Together with comparisons against strong convolutional baselines such as UNet and ResNet, the consistent performance of CineMA demonstrates its robustness and broad applicability. Finally, we recognise that clinical deployment is often limited by computational constraints and the complexity of model development. By releasing all code and model weights, we aim to support reproducibility and accelerate clinical translation through easier evaluation and adaptation of our foundation model across diverse applications.

Nevertheless, there remain several limitations and challenges requiring further exploration. First, although CineMA was evaluated across a broad range of tasks and associations with non-cardiac conditions and survival were examined, we did not explicitly model or predict clinical outcomes such as major adverse cardiovascular events or mortality. Future work could develop outcome-prediction models using CineMA-derived metrics as inputs, to determine whether improved accuracy in functional measurements and disease diagnosis translates into better prognostic performance. Second, CineMA was pre-trained exclusively on cine CMR images from the UK Biobank. Expanding the pre-training dataset to include additional sources could increase population and acquisition diversity, potentially improving generalisability. In addition, we observed limited benefits of CineMA on LGE and T1 mapping images (\Cref{tab:supp_emidec_segmentation}, \Cref{tab:supp_myops2020_segmentation}), which is expected given that these modalities were not included in pre-training. Future work should consider incorporating additional modalities, particularly those from short-axis views. Third, while the proposed convolution–transformer architecture yielded strong performance when fine-tuned, it did not consistently match purely convolutional architectures when trained from random initialisation, suggesting room for further optimisation in architectural design. Finally, training foundation models requires substantial computational resources, particularly due to the large parameter counts in Transformer-based architectures. This presents a practical barrier to adoption in clinical environments. Exploring more efficient architectures, such as mixture-of-experts models~\citep{dai2024deepseekmoe}, could help mitigate this constraint and enhance scalability.

In conclusion, we demonstrated the efficacy and efficiency of CineMA compared to traditional task-specific convolutional models across a range of clinically relevant tasks. Beyond technical performance, we also showed that CineMA-derived functional metrics are clinically meaningful, capturing differences in cardiac function associated with diabetes and cancer, correlating with survival outcomes, and maintaining consistent performance across demographic subgroups. These findings highlight CineMA’s potential for both clinical deployment and large-scale epidemiological research. The public release of the full training and evaluation pipeline, along with model weights, democratises access to rigorous evaluation, modular frameworks, and high-performance models that would otherwise require substantial development effort and computational resources. This will lower the barrier to clinical and commercial adoption, enabling institutions worldwide to benefit from our work and accelerate research in cardiovascular diagnostics.

\section{Methods}
\label{sec:methods}

\subsection{Data for developing CineMA}
We curated 74,916 unlabelled cine CMR studies from the UK Biobank dataset~\citep{raisi2021cardiovascular} using data fields 20208 (LAX heart images) and 20209 (SAX heart images). The LAX images included two-chamber, three-chamber, and four-chamber views. Each study contained 50 images.

\subsection{Data for ventricle segmentation}
We collected 885 SAX cine CMR scans and corresponding segmentation masks of the ventricles and myocardium from three datasets: the Automated Cardiac Diagnosis Challenge (ACDC)~\citep{bernard2018deep}, the Multi-Centre, Multi-Vendor \& Multi-Disease Cardiac Image Segmentation Challenge (M\&Ms)~\citep{campello2021multi}, and M\&Ms2~\citep{martin2023deep}. M\&Ms2 additionally contains LAX 4C views and corresponding segmentation masks. For each subject, only the end-diastolic (ED) and end-systolic (ES) frames were used for training and evaluation. Ventricular and myocardium volumes were calculated as the product of the number of labelled voxels and the voxel volume. The dataset includes both healthy subjects and those with cardiovascular disease.

\subsection{Data for ejection fraction estimation}
For ejection fraction prediction, we derived ventricular and myocardial volumes from the 885 segmentation masks in the ACDC, M\&Ms, and M\&Ms2 datasets. EF was then calculated from the end-diastolic volume (\gls{edv}) and end-systolic volume (\gls{esv}) using the formula:
$$\text{EF} = \frac{\text{EDV} - \text{ESV}}{\text{EDV}} \times 100\%.$$
To assess generalisability, we additionally included 440 and 206 short-axis cine CMR scans and the LVEF labels from the test split of the Kaggle dataset~\citep{second-annual-data-science-bowl} and the Rescan dataset~\citep{bhuva2019multicenter}, respectively. These scans were not used for model training and serve as an external evaluation set. This combined external dataset represents a population-shifted distribution, with a higher average LVEF compared to the other datasets. Since the scans from the Rescan dataset are paired, corresponding to acquisitions at two time points that are within one week in 96\% of cases, their LVEF measurements were averaged to create a shared reference label.

\subsection{Data for cardiovascular disease diagnosis}
For disease classification, we used the ED and ES frames from the ACDC, M\&Ms, and M\&Ms2 datasets, each of which contains a distinct set of diagnostic categories. To create a unified binary classification task, we pooled all disease types into a single ``disease present'' class. This resulted in a binary disease detection dataset comprising all 885 samples.

\subsection{Data for long-axis function estimation}
The landmark dataset, which contains 1,444 studies with LAX 2C and 4C images at both ED and ES phases, was used to estimate long-axis function. For each study, we calculated mitral annular plane systolic excursion (MAPSE) as the average displacement of the mitral valve landmarks from ED to ES:
\begin{align*}
\text{MAPSE} = \frac{1}{2} ( \| \mathbf{p}_1^{\text{ED}} - \mathbf{p}_1^{\text{ES}} \| + \| \mathbf{p}_2^{\text{ED}} - \mathbf{p}_2^{\text{ES}} \| ),
\end{align*}
\noindent
where $\mathbf{p}_1^{\text{ED}}, \mathbf{p}_2^{\text{ED}} \in \mathbb{R}^2 $ are the 2D coordinates of the mitral annular landmarks at ED and ES, and $\| \cdot \|$ denotes the Euclidean norm. For each LAX image, we estimated the ventricular length as the distance between the midpoint of the mitral annular landmarks and the apical landmark:
\begin{align*}
\text{LV Length} = \| \frac{\mathbf{p}_1 + \mathbf{p}_2}{2} - \mathbf{p}_{\text{apex}} \|,
\end{align*}
\noindent
where $\mathbf{p}_{\text{apex}} \in \mathbb{R}^2$ denotes the 2D coordinates of the apical landmark. Finally, global longitudinal shortening (GLS) was estimated in a form analogous to EF:
\begin{align*}
\text{GLS} = \frac{\text{ED LV Length} - \text{ES LV Length}}{\text{ED LV Length}} \times 100\%.
\end{align*}
Additionally, for each image, a heatmap of landmarks was generated using a Gaussian kernel with \( \sigma = 3 \) per landmark:
\begin{align*}
    \text{heatmap}(\mathbf{p}; \mathbf{p}_i) = \exp\left(-\frac{\|\mathbf{p} - \mathbf{p}_i\|^2}{2\sigma^2}\right)
\end{align*}
where \( \mathbf{p} \in \mathbb{R}^2 \) denotes a pixel location and \( \mathbf{p}_i \in \mathbb{R}^2 \) is the ground truth coordinate of the \( i \)-th landmark, with \( i \in \{1, 2, \text{apex} \} \).

\subsection{Data for scar segmentation}
We additionally collected 100 DE-MRI images from the EMIDEC dataset~\citep{lalande2020emidec} and 45 studies from the MyoPS2020 dataset~\citep{li2023myops}, which consist of late gadolinium enhancement (LGE) CMR, T2-weighted CMR, and balanced steady-state free precession (bSSFP) sequences. These are acquired in the short-axis view but exhibit different image intensities compared to standard cine CMRs. Both datasets provide segmentation labels for the ventricular cavity, myocardium, and myocardial scars.

\subsection{Data for diabetes, hypertension, cancer, survival and demographic analysis}
We used the UK Biobank dataset for association analyses with diabetes, cancer, and survival status. The diabetes label was defined using data field 2443 (diabetes diagnosed by doctor). The labels of hypertension and cancer were derived from data field 2966 (age high blood pressure diagnosed) and 40008 (age at cancer diagnosis), respectively. One-year, two-year, and five-year survival labels were derived from data field 40007 (age at death). Since the data was retrieved in 2023, the one-year survival status was determined using records from participants whose last hospital visit occurred in 2022 or earlier; if the age at death was not recorded, the participant was assumed to have survived. The same approach was applied for the two-year and five-year survival definitions. Sex and ethnic group information were obtained from data fields 31 and 21000 (ethnic background), respectively. The LVEF labels were provided by data field 22420 (LV ejection fraction), which was automatically derived from Heart MRI data (inlineVF) without any expert quality control.

\subsection{Data processing}
All image resolutions were standardised to $1\times1\times10~\mathrm{mm}$ for SAX and $1\times1~\mathrm{mm}$ for LAX views. All images were then centre cropped or padded to a fixed spatial size of $192\times192$ for SAX and $256\times256$ for LAX. Centre cropping was performed either at the intersection of SAX, LAX 2C, and LAX 4C views when available, or at the anatomical centre of the left ventricle.

\subsection{CineMA architecture and implementation}
CineMA adopts a masked autoencoder architecture~\citep{he2022masked} that processes four views: two-chamber, three-chamber, and four-chamber LAX views, along with SAX views. The LAX views are two-dimensional with size $(256, 256)$, while the SAX views are three-dimensional with size $(192, 192, 16)$. Each view is first split into patches of size $(16, 16)$, with the SAX view using a depth-wise patch size of 1. A fixed proportion (75\%) of patches is randomly masked by setting them to zero. The input images are then encoded using convolutional downsampling as in ConvMAE~\cite{gao2022convmae}, reducing the spatial dimensions by a factor of $8\times$. The downsampled features are split into non-overlapping $(2, 2)$ patches and embedded into feature tokens. Each masked patch corresponds to a single feature token, allowing the tokens to be categorised into two groups: visible tokens (from unmasked patches) and masked tokens (to be predicted). Following MultiMAE~\citep{bachmann2022multimae}, the visible tokens from all views are concatenated and passed through a shared Vision Transformer encoder. The encoded representations are then combined with the masked tokens and passed through a shared Transformer decoder to reconstruct the masked patches. In this study, CineMA adopted the ``base'' configuration from He et al.~\cite{he2022masked}, resulting in a model with 126 million parameters. It was pre-trained for 800 epochs on 74,916 cine CMR studies from the UK Biobank~\citep{raisi2021cardiovascular}, using LAX 2C, 3C, 4C, and SAX views. For each study, a single image was randomly selected from the available 50 frames. Random gamma augmentation was applied, followed by independent random affine transformations (including rotation, zooming, shearing, and shifting) to each view. The model was trained with a batch size of 128. The learning rate schedule involved a linear warm-up over the first 10 epochs from 0 to 0.001, followed by a cosine decay to $10^{-6}$. Training used a mean squared error loss on masked patches and the AdamW optimiser with betas (0.9, 0.95). Gradients were clipped with a maximum norm of 5.0. The final checkpoint was used for all downstream tasks.

\subsection{Adaptation to downstream tasks}

In downstream task fine-tuning, we retained only the encoder branches corresponding to the available views and discarded all decoder layers. Additional layers were added with randomly initialised weights depending on the type of label. A UNetR-style architecture~\citep{hatamizadeh2022unetr} was employed for segmentation masks and landmark location heatmaps. For tabular labels, such as disease classification or EF values, linear prediction layers were added. During fine-tuning, all model weights were learnable, and a weight decay of 0.05 was applied to the Transformer layers. Data augmentation techniques included random gamma adjustment, random scaling, and random affine transformations for all tasks. Label smoothing was applied during classification training, and random image dropout was used during segmentation training. Early stopping was employed based on validation performance. All models used the same set of hyperparameters, listed in \Cref{tab:fine-tune_hparam}, which were fixed across experiments and not tuned via exhaustive search.

\subsection{Convolutional neural network baselines}
For dense prediction tasks such as segmentation and heatmap regression, five-layer UNet models with residual connections were employed. These models used channel widths of 32, 64, 128, 256, and 512 across successive layers. For tabular labels, such as EF estimation and disease classification, ResNet50~\citep{targ2016resnet} was used as baseline models. 2D and 3D models were used for LAX and SAX images, respectively. All convolutional neural networks were trained using the same procedures as those used for fine-tuning CineMA. The nnUNet~\citep{isensee2021nnu} was not included in benchmarks due to its automated optimisation pipeline. Fair comparison with such an AutoML optimisation to foundation models remains an open research direction.

\subsection{Downstream task evaluations}
Segmentation performance was assessed using Dice score and 95th percentile Hausdorff distance. Disease classification tasks were evaluated using the area under the receiver operating characteristic curve (ROC AUC), and for disease detection, sensitivity and specificity were reported. Labels with continuous values such as volumes, EF, MAPSE, and GLS were evaluated using mean absolute error. To evaluate model consistency across repeated acquisitions in Rescan dataset, the coefficient of variation was calculated. Landmark localisation accuracy was evaluated using Euclidean distance. For each task, we trained the model three times using different random seeds, either on the full dataset or on a randomly sampled subset. When subsets were used, the training samples varied across seeds, while the validation set remained fixed. We applied bootstrapping on each test set by sampling $N$ examples with replacement ($N$ = test set size). Metrics were averaged across the three trained model checkpoints for each bootstrap sample, and this process was repeated 100 times to compute statistics such as mean and standard deviation. Statistical significance was evaluated using two-sided t-tests on the bootstrapped metric values, at varying significance levels, denoted *** for $p < 0.001$, ** for $p < 0.01$, and * for $p < 0.05$; ns indicates non-significance.

\subsection{Association analysis}
The association analyses with diabetes, hypertension, and cancer were performed using the ordinary least squares linear model from \texttt{statsmodels}, where the cardiac functional metrics (such as LVEF) were predicted using disease presence, age, sex, and BMI:
\begin{align*}
    y_\text{LVEF} = \text{constant} + \beta_1x_\text{disease} + \beta_2x_\text{age} + \beta_3x_\text{sex} + \beta_4x_\text{BMI}
\end{align*}
The survival analysis was performed by fitting a Cox’s proportional hazard model from \texttt{lifelines}:
\begin{align*}
h(t) = h_0(t) \exp(\beta_1x_\text{disease} + \beta_2x_\text{age} + \beta_3x_\text{sex} + \beta_4x_\text{BMI})
\end{align*}

\subsection{Demographic disparity analysis}
For each cardiac function metric (denoted $m$), the disparity ratio at a given percentile $q$ is defined as the ratio of positive outcome ($m\geq q$) rates between white and non-white subjects,
\begin{align*}
\text{Disparity Ratio} = \frac{n_{\text{White},m>q}}{n_\text{White}} / \frac{n_{\text{Non-white},m>q}}{n_\text{Non-white}}.
\end{align*}

\subsection{Computational resources}
The pre-training of CineMA was completed in 18 days using 8 NVIDIA RTX A6000 GPUs on the University College London Computer Science HPC cluster. Downstream training and fine-tuning were conducted on both the University College London Computer Science HPC cluster and the Isambard-AI cluster~\citep{mcintosh2024isambard}. The duration of downstream training ranged from several hours to a maximum of one day, depending on dataset size and GPU availability.

\section{Data availability}
The UK Biobank cardiac MRI data can be accessed by bona fide researchers through an application to the UK Biobank (\url{https://www.ukbiobank.ac.uk}). The ACDC dataset is available at \url{https://www.creatis.insa-lyon.fr/Challenge/acdc/databases.html}. The M\&Ms and M\&Ms-2 datasets can be requested at \url{https://www.ub.edu/mnms} and \url{https://www.ub.edu/mnms-2/}. The Kaggle cardiac MRI dataset is available at \url{https://www.kaggle.com/c/second-annual-data-science-bowl/data}. The EMIDEC dataset is available at \url{https://emidec.com}. The MyoPS2020 dataset is available at \url{https://zmiclab.github.io/zxh/0/myops20/}. 
The Landmark dataset is not publicly available, but investigators from research institutions may request access by contacting the authors directly.
Rescan data is available at \url{https://thevolumesresource.com/}.

\section{Code availability}
All code for pre-training, fine-tuning, and evaluation, along with weights of pre-trained and fine-tuned models, are publicly available under an MIT license at \url{https://github.com/mathpluscode/CineMA}.

\section{Acknowledgement}

This research has been conducted using the UK Biobank Resource under Application Number 71702. The authors acknowledge the use of resources provided by the Isambard-AI National AI Research Resource (AIRR). Isambard-AI is operated by the University of Bristol and is funded by the UK Government’s Department for Science, Innovation and Technology (DSIT) via UK Research and Innovation; and the Science and Technology Facilities Council [ST/AIRR/I-A-I/1023].
CM, JM, RD and TT are directly and indirectly supported by the NIHR Biomedical Research Centres at University College London Hospital and Barts Health NHS Trusts. JM and CM are co-founders of Mycardium.

\section{Declaration of interest}
The authors declare no competing interests.

\section{Extended data figures and tables}

\begin{table}[!htbp]
\centering
\caption{Datasets used in pre-training and downstream tasks. n is the number of subjects in each dataset. Each dataset includes different views and may contain a different number of frames. SAX, short-axis. LAX, long-axis. nC, n-chamber. EF, ejection fraction. BMI, body mass index.}
\label{tab:data_labels}
\begin{tabular}{lllll}
\toprule
\textbf{Dataset} & \textbf{Modality} & \textbf{Views} & \textbf{n} & \textbf{Tasks}\\ \midrule
UKB~\citep{raisi2021cardiovascular} & Cine & \makecell[l]{SAX;\\LAX 2C/3C/4C} & 74,916 & \makecell[l]{Pre-training;\\Diabetes, hypertension, cancer;\\1-/2-/5-year survival.}\\ \midrule
ACDC~\citep{bernard2018deep} & Cine & SAX & 150 & \makecell[l]{Ventricle segmentation;\\EF and BMI regression;\\Disease classification.}\\ \midrule 
M\&Ms~\citep{campello2021multi} & Cine & SAX & 375 & \makecell[l]{Ventricle segmentation;\\EF and age regression;\\Disease and sex classification.}\\ \midrule 
M\&Ms2~\citep{martin2023deep} & Cine & \makecell[l]{SAX;\\LAX 4C} & 360 & \makecell[l]{Ventricle segmentation;\\EF regression;\\Disease and vendor classification.}\\ \midrule 
Kaggle~\citep{second-annual-data-science-bowl} & Cine & SAX & 440 & EF regression. \\ \midrule 
Rescan~\citep{bhuva2019multicenter} & Cine & SAX & 206 & EF regression.\\ \midrule 
Landmark~\citep{xue2021landmark} & Cine & LAX 2C/4C & 2,888 & \makecell[l]{Landmark localisation;\\Long-axis function regression.}\\ \midrule 
EMIDEC~\citep{lalande2020emidec} & LGE & SAX & 100 & Ventricle segmentation.\\ \midrule 
Myops2020~\citep{li2023myops} & \makecell[l]{LGE,\\T2,\\bSSFP} & SAX & 25 & Scar segmentation.\\ 
\bottomrule
\end{tabular}
\end{table}

\begin{table}[!htbp]
\centering
\begin{tabular}{ll|cccc|cccc}
\toprule
\multirow{2}{*}{Dataset} & \multirow{2}{*}{Model} & \multicolumn{4}{c}{Dice Score (\%)} & \multicolumn{4}{c}{95\% Hausdorff Distance (mm)} \\
 &  & RV & MYO & LV & Mean & RV & MYO & LV & Mean \\
\midrule
\multirow{4}{*}{\makecell[l]{ACDC\\(SAX)}} & UNet\textsuperscript{RandInit} & 89.33 & 86.71 & 92.58 & 89.54 & 5.41 & 3.93 & 3.86 & 4.40 \\
& CineMA\textsuperscript{RandInit} & 90.14 & 86.51 & 91.98 & 89.54 & 5.41 & 4.64 & 4.92 & 5.01 \\
& CineMA\textsuperscript{FineTune} & 91.06 & 87.93 & 93.56 & 90.85 & 4.29 & 3.02 & 3.04 & 3.45 \\
& Jacob et al.~\citep{jacob2024towards} & 90.7 & 87.9 & 93.3 & 90.63 & \\ \hline
\multirow{3}{*}{\makecell[l]{M\&Ms\\(SAX)}} & UNet\textsuperscript{RandInit} & 86.14 & 82.09 & 89.66 & 85.96 & 6.75 & 6.06 & 6.31 & 6.37 \\
 & CineMA\textsuperscript{RandInit} & 84.65 & 81.12 & 89.23 & 85.00 & 10.26 & 6.25 & 6.80 & 7.77 \\
& CineMA\textsuperscript{FineTune} & 87.80 & 83.35 & 90.70 & 87.28 & 5.81 & 5.15 & 5.00 & 5.32 \\ \hline
\multirow{3}{*}{\makecell[l]{M\&Ms2\\(SAX)}} & UNet\textsuperscript{RandInit} & 88.26 & 83.56 & 92.25 & 88.02 & 8.63 & 6.49 & 6.37 & 7.16 \\
& CineMA\textsuperscript{RandInit} & 84.99 & 81.91 & 91.02 & 85.97 & 14.08 & 7.44 & 7.39 & 9.69 \\
& CineMA\textsuperscript{FineTune} & 89.72 & 84.80 & 93.12 & 89.21 & 7.17 & 5.31 & 5.27 & 5.91 \\ \hline
\multirow{3}{*}{\makecell[l]{M\&Ms2\\(LAX 4C)}}& UNet\textsuperscript{RandInit} & 90.48 & 86.69 & 94.64 & 90.60 & 5.96 & 4.37 & 4.28 & 4.87 \\
& CineMA\textsuperscript{RandInit} & 89.83 & 86.81 & 94.72 & 90.45 & 6.75 & 4.52 & 4.26 & 5.17 \\
& CineMA\textsuperscript{FineTune} & 91.09 & 87.64 & 95.59 & 91.44 & 5.22 & 3.61 & 3.39 & 4.07 \\ \midrule
\multirow{2}{*}{Dataset} & \multirow{2}{*}{Model} & \multicolumn{4}{c}{Volume MAE (ml)} & \\
 &  & RV & MYO & LV & Mean \\
\midrule
\multirow{3}{*}{\makecell[l]{ACDC\\(SAX)}} & UNet\textsuperscript{RandInit} & 11.29 & 7.98 & 5.73 & 8.33 \\
 & CineMA\textsuperscript{RandInit} & 8.89 & 8.13 & 6.16 & 7.72 \\
 & CineMA\textsuperscript{FineTune} & 9.30 & 7.53 & 5.07 & 7.30 \\ \hline
\multirow{3}{*}{\makecell[l]{M\&Ms\\(SAX)}} & UNet\textsuperscript{RandInit} & 10.47 & 13.71 & 9.42 & 11.20 \\
 & CineMA\textsuperscript{RandInit} & 12.60 & 12.92 & 8.71 & 11.41 \\
 & CineMA\textsuperscript{FineTune} & 9.29 & 11.57 & 8.54 & 9.80 \\ \hline
\multirow{3}{*}{\makecell[l]{M\&Ms2\\(SAX)}} & UNet\textsuperscript{RandInit} & 12.75 & 8.15 & 8.10 & 9.67 \\
 & CineMA\textsuperscript{RandInit} & 17.47 & 8.66 & 9.00 & 11.71 \\
 & CineMA\textsuperscript{FineTune} & 12.62 & 7.33 & 7.39 & 9.12 \\
\bottomrule
\end{tabular}
\caption{Ventricle and myocardium segmentation performance. Fine-tuned CineMA outperformed baselines across all datasets. It also outperformed Jacob et al.~\citep{jacob2024towards} on ACDC. RV, right ventricle. MYO, myocaridum. LV, left ventricle. MAE, mean absolute error.}
\label{tab:supp_cine_segmentation}
\end{table}

\begin{table}[!htbp]
\centering
\begin{tabular}{ll|cc}
\toprule
Dataset & Model & LVEF MAE (\%) & RVEF MAE (\%)\\
\midrule
\multirow{3}{*}{\makecell[l]{ACDC\\(SAX)}} & UNet\textsuperscript{RandInit} & 2.50 & 5.48 \\
& CineMA\textsuperscript{RandInit} & 2.99 & 4.54 \\
& CineMA\textsuperscript{FineTune} & 2.26 & 4.08 \\ \hline
\multirow{3}{*}{\makecell[l]{M\&Ms\\(SAX)}} & UNet\textsuperscript{RandInit} & 4.89 & 6.38 \\
& CineMA\textsuperscript{RandInit} & 4.71 & 7.11 \\
& CineMA\textsuperscript{FineTune} & 4.20 & 6.19 \\ \hline
\multirow{3}{*}{\makecell[l]{M\&Ms2\\(SAX)}} & UNet\textsuperscript{RandInit} & 3.50 & 6.97 \\
& CineMA\textsuperscript{RandInit} & 4.24 & 8.70 \\
& CineMA\textsuperscript{FineTune} & 3.53 & 6.20 \\ \hline
\multirow{3}{*}{\makecell[l]{M\&Ms2\\(LAX 4C)}} & UNet\textsuperscript{RandInit} & 3.21 & 6.09 \\
& CineMA\textsuperscript{RandInit} & 3.17 & 6.61 \\
& CineMA\textsuperscript{FineTune} & 3.07 & 5.59 \\
\bottomrule
\end{tabular}
\caption{Ejection fraction (EF) regression performance using segmentation models. Fine-tuned CineMA outperformed baselines across all datasets. RV, right ventricle. LV, left ventricle. MAE, mean absolute error.}
\label{tab:supp_cine_segmentation_ef}
\end{table}

\begin{table}[!htbp]
\centering
\begin{tabular}{ll|c|cc}
\toprule
\multirow{2}{*}{Dataset} & \multirow{2}{*}{Model} & Kaggle & \multicolumn{2}{c}{Rescan} \\
 & & MAE (\%) & MAE (\%) & CV (\%)\\
\midrule
ACDC & UNet\textsuperscript{RandInit} & 6.24 & 9.49 & 19.01 \\
 & CineMA\textsuperscript{RandInit} & 6.03 & 7.36 & 11.38 \\
 & CineMA\textsuperscript{FineTune} & 5.21 & 7.98 & 10.70 \\ \hline
M\&Ms & UNet\textsuperscript{RandInit} & 5.08 & 3.90 & 7.65 \\
& CineMA\textsuperscript{RandInit} & 5.98 & 4.18 & 7.82 \\
& CineMA\textsuperscript{FineTune} & 3.87 & 4.08 & 7.84 \\ \hline
M\&Ms2 & UNet\textsuperscript{RandInit} & 4.18 & 4.86 & 9.39 \\
& CineMA\textsuperscript{RandInit} & 5.42 & 4.92 & 11.27 \\
& CineMA\textsuperscript{FineTune} & 3.77 & 4.05 & 6.82 \\ \hline
UKB & Shad et al.~\citep{shad2023generalizable} & 6.88 & \\ \hline
Rescan & Bhuva et al.~\citep{bhuva2019multicenter} &  & 4 & 8.8\\ 
\bottomrule
\end{tabular}
\caption{Segmentation models' zero-shot performance on left ventricle ejection fraction regression using short-axis images. Models were trained on ACDC, M\&Ms, and M\&Ms2, respectively, then evaluated on Kaggle and Rescan. Fine-tuned CineMA outperformed baselines and previous methods on each dataset. MAE, mean absolute error. CV, coefficient of variation.}
\label{tab:supp_cine_segmentation_ef_ood}
\end{table}

\begin{table}[!htbp]
\centering
\begin{tabular}{ll|cc|cc}
\toprule
\multirow{2}{*}{Data} & \multirow{2}{*}{Model} & \multicolumn{2}{c}{Classification} & \multicolumn{2}{c}{Detection} \\
 &  & ROC AUC (\%) & F1 (\%) & Specificity (\%) & Sensitivity (\%)\\
\midrule
\multirow{3}{*}{\makecell[l]{ACDC\\(SAX)}} 
& ResNet\textsuperscript{RandInit} & 81.05 & 50.26 & 44.59 & 82.81 \\
& CineMA\textsuperscript{RandInit} & 63.13 & 14.43 & 0.00 & 99.16 \\
& CineMA\textsuperscript{FineTune} & 97.98 & 84.27 & 83.03 & 94.87 \\
\hline
\multirow{3}{*}{\makecell[l]{M\&Ms\\(SAX)}} 
& ResNet\textsuperscript{RandInit} & 63.34 & 26.49 & 27.40 & 85.43 \\
& CineMA\textsuperscript{RandInit} & 58.24 & 16.54 & 5.92 & 93.96 \\
& CineMA\textsuperscript{FineTune} & 77.41 & 39.92 & 66.20 & 84.05 \\
\hline
\multirow{3}{*}{\makecell[l]{M\&Ms2\\(SAX)}} 
& ResNet\textsuperscript{RandInit} & 69.09 & 35.31 & 22.66 & 92.29 \\
& CineMA\textsuperscript{RandInit} & 61.58 & 23.66 & 29.12 & 77.13 \\
& CineMA\textsuperscript{FineTune} & 77.20 & 47.70 & 30.96 & 91.69 \\
\hline
\multirow{3}{*}{\makecell[l]{M\&Ms2\\(LAX 4C)}} 
& ResNet\textsuperscript{RandInit} & 71.87 & 35.88 & 40.35 & 78.31 \\
& CineMA\textsuperscript{RandInit} & 64.90 & 23.73 & 32.16 & 74.65 \\
& CineMA\textsuperscript{FineTune} & 80.86 & 46.31 & 31.73 & 92.02 \\
\bottomrule
\end{tabular}
\caption{Cardiovascular disease classification performance on different training datasets. Fine-tuned CineMA outperformed baselines in most cases. ROC AUC, area under the receiver operating characteristic curve.}
\label{tab:supp_cvd_classification}
\end{table}

\begin{table}[!htbp]
\centering

\begin{tabular}{l|ll|cc}
\toprule
Target & Data & Model & ROC AUC (\%) & F1 (\%) \\
\midrule
\multirow{6}{*}{Vendor} 
& \multirow{3}{*}{\makecell[l]{M\&Ms2\\(SAX)}} 
& ResNet\textsuperscript{RandInit} & 79.64 & 55.57 \\
&  & CineMA\textsuperscript{RandInit} & 80.25 & 58.68 \\
&  & CineMA\textsuperscript{FineTune} & 81.10 & 62.52 \\
& \multirow{3}{*}{\makecell[l]{M\&Ms2\\(LAX 4C)}} 
& ResNet\textsuperscript{RandInit} & 84.41 & 63.32 \\
&  & CineMA\textsuperscript{RandInit} & 81.29 & 59.53 \\
&  & CineMA\textsuperscript{FineTune} & 77.76 & 63.12 \\
\hline
\multirow{3}{*}{Sex} 
& \multirow{3}{*}{\makecell[l]{M\&Ms\\(SAX)}} 
& ResNet\textsuperscript{RandInit} & 86.17 & 70.24 \\
&  & CineMA\textsuperscript{RandInit} & 71.08 & 32.44 \\
&  & CineMA\textsuperscript{FineTune} & 96.90 & 90.75 \\
\bottomrule
\end{tabular}
\caption{Direct image classification performance on vendor and sex. CineMA\textsuperscript{FineTune} outperformed CineMA\textsuperscript{RandInit}. It also outperformed or matched UNet. ROC AUC, area under the receiver operating characteristic curve.}
\label{tab:supp_vendor_sex_classification}
\end{table}

\begin{table}[!htbp]
\centering

\begin{tabular}{l|ll|r}
\toprule
Target & Data & Model & MAE \\
\midrule
\multirow{12}{*}{EF} & \multirow{3}{*}{\makecell[l]{ACDC\\(SAX)}} & ResNet\textsuperscript{RandInit} & 5.58 \\
 & & CineMA\textsuperscript{RandInit} & 11.01 \\
 & & CineMA\textsuperscript{FineTune} & 5.07 \\
 & \multirow{3}{*}{\makecell[l]{M\&Ms\\(SAX)}} & ResNet\textsuperscript{RandInit} & 6.94 \\
 & & CineMA\textsuperscript{RandInit} & 7.33 \\
 &  & CineMA\textsuperscript{FineTune} & 5.41 \\
 & \multirow{3}{*}{\makecell[l]{M\&Ms2\\(SAX)}} & ResNet\textsuperscript{RandInit} & 8.19 \\
 & & CineMA\textsuperscript{RandInit} & 7.13 \\
 & & CineMA\textsuperscript{FineTune} & 5.15 \\
 & \multirow{3}{*}{\makecell[l]{M\&Ms2\\(LAX 4C)}} & ResNet\textsuperscript{RandInit} & 7.42 \\
 & & CineMA\textsuperscript{RandInit} & 7.60 \\
 & & CineMA\textsuperscript{FineTune} & 6.70 \\ \hline
\multirow{3}{*}{BMI} & \multirow{3}{*}{\makecell[l]{ACDC\\(SAX)}} & ResNet\textsuperscript{RandInit} & 4.36 \\
 & & CineMA\textsuperscript{RandInit} & 4.17 \\
 & & CineMA\textsuperscript{FineTune} & 3.09 \\ \hline
\multirow{3}{*}{Age} & \multirow{3}{*}{\makecell[l]{M\&Ms\\(SAX)}} & ResNet\textsuperscript{RandInit} & 12.38 \\
 &  & CineMA\textsuperscript{RandInit} & 12.51 \\
 &  & CineMA\textsuperscript{FineTune} & 8.68 \\
\bottomrule
\end{tabular}
\caption{Direct image regression performance on ejection fraction (EF), body mass index (BMI) and age. Fine-tuned CineMA outperformed baselines consistently. MAE, mean absolute error.}
\label{tab:supp_regression}
\end{table}

\begin{table}[!htbp]
\centering
\begin{tabular}{ll|cccc|cc}
\toprule
\multirow{2}{*}{View} & \multirow{2}{*}{Model} & \multicolumn{4}{c}{Euclidean Distance (mm)} & \multicolumn{2}{c}{MAE (\%)} \\
 &  & Valve1 & Valve2 & Apex & Mean & MAPSE & GLS \\
\midrule
\multirow{4}{*}{LAX 2C} & UNet\textsuperscript{RandInit} & 0.85 & 0.90 & 0.93 & 0.89 & 0.56 & 1.22 \\
& CineMA\textsuperscript{RandInit} & 0.93 & 1.00 & 1.10 & 1.01 & 0.67 & 1.37 \\
& CineMA\textsuperscript{FineTune} & 0.83 & 0.89 & 0.99 & 0.90 & 0.61 & 1.24 \\
& Xue et al.~\citep{xue2021landmark} & 2.1 & 2.1 & 2.4 & 2.20 &  &  \\ \hline
\multirow{4}{*}{LAX 4C} & UNet\textsuperscript{RandInit} & 0.98 & 1.25 & 1.07 & 1.10 & 0.69 & 1.35 \\
& CineMA\textsuperscript{RandInit} & 1.13 & 1.31 & 1.26 & 1.23 & 0.77 & 1.55 \\
 & CineMA\textsuperscript{FineTune} & 0.99 & 1.24 & 1.06 & 1.10 & 0.71 & 1.32 \\
 & Xue et al.~\citep{xue2021landmark} & 3.4 & 2.1 & 2.8 & 2.77 &  &  \\ 
\bottomrule
\end{tabular}
\caption{Landmark localisation and long-axis function performance using heatmap regression. CineMA\textsuperscript{FineTune} outperformed CineMA\textsuperscript{RandInit} and matched UNet. It also outperformed previous method. MAE, mean absolute error.}
\label{tab:supp_landmark}
\end{table}

\begin{table}[!htbp]
\centering

\begin{tabular}{l|ll|cccc|cc}
\toprule
\multirow{2}{*}{Method} & \multirow{2}{*}{View} & \multirow{2}{*}{Model} & \multicolumn{4}{c}{Euclidean Distance (mm)} & \multicolumn{2}{c}{MAE (\%)} \\
 & &  & Valve1 & Valve2 & Apex & Mean & MAPSE & GLS \\
\midrule
\multirow{6}{*}{Heatmap} & \multirow{3}{*}{LAX 2C} & UNet\textsuperscript{RandInit} & 0.85 & 0.90 & 0.93 & 0.89 & 0.56 & 1.22 \\
& & CineMA\textsuperscript{RandInit} & 0.93 & 1.00 & 1.10 & 1.01 & 0.67 & 1.37 \\
& & CineMA & 0.83 & 0.89 & 0.99 & 0.90 & 0.61 & 1.24 \\
& \multirow{3}{*}{LAX 4C} & UNet\textsuperscript{RandInit} & 0.98 & 1.25 & 1.07 & 1.10 & 0.69 & 1.35 \\
& & CineMA\textsuperscript{RandInit} & 1.13 & 1.31 & 1.26 & 1.23 & 0.77 & 1.55 \\
&  & CineMA & 0.99 & 1.24 & 1.06 & 1.10 & 0.71 & 1.32 \\ \hline
\multirow{6}{*}{Coordinates} & \multirow{3}{*}{LAX 2C} & ResNet\textsuperscript{RandInit} & 1.88 & 1.88 & 1.87 & 1.88 & 0.87 & 1.66 \\
& & CineMA\textsuperscript{RandInit} & 7.96 & 7.90 & 8.37 & 8.07 & 4.49 & 6.41 \\
& & CineMA & 1.06 & 1.13 & 1.09 & 1.09 & 0.66 & 1.24 \\
& \multirow{3}{*}{LAX 4C} & ResNet\textsuperscript{RandInit} & 1.99 & 4.18 & 4.60 & 3.59 & 1.01 & 1.50 \\
& & CineMA\textsuperscript{RandInit} & 5.61 & 5.48 & 5.78 & 5.62 & 1.97 & 3.22 \\
& & CineMA & 1.18 & 1.39 & 1.14 & 1.24 & 0.75 & 1.30 \\
\bottomrule
\end{tabular}
\caption{Landmark localisation performance using heatmap or coordinates regression. Coordinate-based model performance was generally lower than with heatmap-based regression. Given its accuracy and interpretability, heatmap regression remains the preferred approach for landmark localisation and long-axis function estimation.}
\label{tab:supp_landmark_compare}
\end{table}

\begin{table}[!htbp]
\centering

\begin{tabular}{l|ccccc|ccc}
\toprule
\multirow{2}{*}{Model} & \multicolumn{5}{c}{Dice Score (\%)} & \multicolumn{3}{c}{95\% Hausdorff Distance (mm)} \\
 & Cavity & MYO & MI & NR & Mean & Cavity & MYO & Mean \\
\midrule
UNet\textsuperscript{RandInit} & 92.83 & 81.94 & 36.55 & 68.16 & 69.87 & 2.94 & 3.41 & 3.18 \\
CineMA\textsuperscript{RandInit} & 75.69 & 51.41 & 7.45 & 54.52 & 47.27 & 10.08 & 11.90 & 10.99 \\
CineMA\textsuperscript{FineTune} & 92.76 & 82.48 & 38.36 & 70.99 & 71.15 & 2.90 & 3.42 & 3.16 \\
\bottomrule
\end{tabular}
\caption{Segmentation performance on EMIDEC. CineMA\textsuperscript{FineTune} consistently outperformed the CineMA\textsuperscript{RandInit} and matched UNet. MYO, myocardium. MI, myocardium infarction. NR, no-reflow. }
\label{tab:supp_emidec_segmentation}
\end{table}

\begin{table}[!htbp]
\centering

\begin{tabular}{l|ccc}
\toprule
\multirow{2}{*}{Model} & \multicolumn{3}{c}{Dice Score (\%)} \\
& Scar & Edema Scar & Mean \\
\midrule
UNet\textsuperscript{RandInit} & 55.24 & 58.74 & 56.99 \\
CineMA\textsuperscript{RandInit} & 52.45 & 59.82 & 56.14 \\
CineMA\textsuperscript{FineTune} & 52.98 & 60.35 & 56.67 \\
\bottomrule
\end{tabular}
\caption{Segmentation performance on Myops2020. CineMA\textsuperscript{FineTune} consistently outperformed the CineMA\textsuperscript{RandInit} and matched UNet.}
\label{tab:supp_myops2020_segmentation}
\end{table}

\begin{table}[!htbp]
\centering
\begin{tabular}{ll|ccc}
\toprule
View & Metric & Coefficient & Confidence Interval & p-value \\
\midrule
SAX & LVEF (UKB)       & -1.3305 & [-1.707, -0.954] & 0.000 \\
SAX & LVEF     & -0.4114 & [-0.671, -0.152] & 0.002 \\
SAX & RVEF     & -0.6916 & [-0.985, -0.399] & 0.000 \\
LAX 2C & GLS           & -0.2722 & [-0.563,  0.019] & 0.067 \\
LAX 4C & GLS           & -0.1750 & [-0.578,  0.228] & 0.395 \\
LAX 2C & MAPSE         & -1.0115 & [-1.165, -0.858] & 0.000 \\
LAX 4C & MAPSE         & -0.5695 & [-0.692, -0.447] & 0.000 \\

\bottomrule
\end{tabular}
\caption{Association between cardiac function and diabetes.}
\label{tab:diabetes}
\end{table}

\begin{table}[!htbp]
\centering
\begin{tabular}{ll|ccc}
\toprule
View & Metric & Coefficient & Confidence Interval & p-value \\
\midrule
SAX & LVEF (UKB)       & 0.7514  & [0.600, 0.902]   & 0.000 \\
SAX & LVEF     & 0.5851  & [0.484, 0.686]   & 0.000 \\
SAX & RVEF     & 0.9049  & [0.791, 1.019]   & 0.000 \\
LAX 2C & GLS           & -0.1630 & [-0.276, -0.050] & 0.005 \\
LAX 4C & GLS           & 0.1231  & [-0.034, 0.280]  & 0.124 \\
LAX 2C & MAPSE         & -0.1413 & [-0.201, -0.082] & 0.000 \\
LAX 4C & MAPSE         & 0.0388  & [-0.009, 0.087]  & 0.111 \\
\bottomrule
\end{tabular}
\caption{Association between cardiac function and hypertension (high blood pressure).}
\label{tab:blood_pressure}
\end{table}

\begin{table}[!htbp]
\centering
\begin{tabular}{ll|ccc}
\toprule
View & Metric & Coefficient & Confidence Interval & p-value \\
\midrule
SAX & LVEF (UKB)       & -0.3440 & [-0.510, -0.178] & 0.000 \\
SAX & LVEF     & -0.3877 & [-0.501, -0.274] & 0.000 \\
SAX & RVEF     & -0.2783 & [-0.407, -0.150] & 0.000 \\
LAX 2C & GLS           & -0.0502 & [-0.178,  0.077] & 0.441 \\
LAX 4C & GLS           & -0.1748 & [-0.352,  0.002] & 0.053 \\
LAX 2C & MAPSE         & -0.1172 & [-0.185, -0.050] & 0.001 \\
LAX 4C & MAPSE         & -0.1349 & [-0.189, -0.081] & 0.000 \\
\bottomrule
\end{tabular}
\caption{Association between cardiac function and cancer.}
\label{tab:cancer}
\end{table}

\begin{table}[!htbp]
\centering
\begin{tabular}{l|ll|ccc}
\toprule
Event & View & Metric & Effect Size & Confidence Interval & p-value \\
\midrule
\multirow{5}{*}{One-year survival} 
& SAX & LVEF (UKB) & -0.04 & [-0.07, -0.01] & 0.02 \\
& SAX & LVEF & -0.05 & [-0.09, -0.01] & 0.01 \\
& SAX & RVEF & 0.02 & [-0.03, 0.06] & 0.44 \\
& LAX 2C & GLS & 0.00 & [-0.04, 0.04] & 0.98 \\
& LAX 4C & GLS & 0.01 & [-0.01, 0.03] & 0.49 \\
& LAX 2C & MAPSE & -0.05 & [-0.13, 0.04] & 0.28 \\
& LAX 4C & MAPSE & -0.05 & [-0.15, 0.05] & 0.34 \\ \midrule
\multirow{5}{*}{Two-year survival} 
& SAX & LVEF (UKB) & -0.03 & [-0.05, -0.02] & $<0.005$ \\
& SAX & LVEF & -0.03 & [-0.05, -0.01] & $<0.005$ \\
& SAX & RVEF & 0.00 & [-0.03, 0.02] & 0.70 \\
& LAX 2C & GLS & 0.00 & [-0.01, 0.02] & 0.70 \\
& LAX 4C & GLS & 0.00 & [-0.02, 0.01] & 0.86 \\
& LAX 2C & MAPSE & -0.04 & [-0.08, 0.01] & 0.11 \\
& LAX 4C & MAPSE & -0.03 & [-0.09, 0.02] & 0.23 \\ \midrule
\multirow{5}{*}{Five-year survival} 
& SAX & LVEF (UKB) & -0.03 & [-0.04, -0.02] & $<0.005$ \\
& SAX & LVEF & -0.03 & [-0.04, -0.01] & $<0.005$ \\
& SAX & RVEF & -0.01 & [-0.03, 0.00] & $0.03$ \\
& LAX 2C & GLS & -0.01 & [-0.03, 0.00] & 0.16 \\
& LAX 4C & GLS & 0.00 & [-0.01, 0.01] & 0.98 \\
& LAX 2C & MAPSE & -0.05 & [-0.08, -0.03] & $<0.005$ \\
& LAX 4C & MAPSE & -0.04 & [-0.08, -0.01] & 0.01 \\
\bottomrule
\end{tabular}
\caption{Survival analysis using Cox's proportional hazard model with cardiac functional metrics.}
\label{tab:survival}
\end{table}

\begin{table}
    \centering
    \caption{Hyperparameters for downstream task model training. If the training dataset size is smaller than the preset batch size, the batch size is reduced to the closest power of two.}
    \label{tab:fine-tune_hparam}
    \begin{tabular}{l|ccc}
        \toprule
        \textbf{Hyperparameter} & \textbf{Classification} & \textbf{Regression} & \textbf{Segmentation}\\ \midrule
        Total Epochs & 800 & 800 & 4000 \\ 
        Warmup Epochs & 10 & 10 & 50 \\ 
        Peak Learning Rate & 0.001 & 0.001 & 0.001\\ 
        End Learning Rate & $10^{-5}$ & $10^{-5}$ & $10^{-5}$\\ 
        Batch Size & 64 & 64 & 64 \\
        Loss & CE Loss & MSE & Dice Loss + CE Loss \\ \midrule
        Validation Frequency & 20 & 20 & 100 \\
        Validation Patience & 5 & 5 & 5\\
        Validation Metric & MCC & Absolute Error & Dice Score\\
        \midrule
        \multirow{2}{*}{\textbf{Hyperparameter}}
         & \multicolumn{2}{c}{\textbf{Landmark localisation}} \\
         & \textbf{Heatmap} & \textbf{Coordinate} \\ \midrule
        Total Epochs & 400 & 400 \\ 
        Warmup Epochs & 10 & 10 \\ 
        Peak Learning Rate & 0.001 & 0.001\\ 
        End Learning Rate & $10^{-5}$ & $10^{-5}$\\ 
        Batch Size & 64 & 64 \\
        Loss & Dice Loss + CE Loss & Wing Loss~\citep{feng2018wing} \\ \midrule
        Validation Frequency & 20 & 20 \\
        Validation Patience & 5 & 5\\
        Validation Metric & $L^2$ distance & $L^2$ distance \\ \bottomrule
    \end{tabular}
\end{table}

\clearpage
\newpage
\begin{appendices}
\setlength{\glsdescwidth}{0.8\linewidth}
\renewcommand{\glsnamefont}[1]{\textbf{#1}}
\printglossary[nogroupskip=true,type=\acronymtype,style=long,nonumberlist]




\end{appendices}


\clearpage
\newpage
\bibliography{sn-bibliography}

\begin{thebibliography}{10}
\expandafter\ifx\csname url\endcsname\relax
  \def\url#1{\burl{#1}}\fi
\expandafter\ifx\csname urlprefix\endcsname\relax\def\urlprefix{URL }\fi
\providecommand{\bibinfo}[2]{#2}
\providecommand{\eprint}[2][]{\url{#2}}
\providecommand{\doi}[1]{\url{https://doi.org/#1}}
\bibcommenthead

\bibitem{mensah2023heart}
\bibinfo{author}{Mensah, G.~A.}, \bibinfo{author}{Fuster, V.} \& \bibinfo{author}{Roth, G.~A.}
\newblock \bibinfo{title}{A heart-healthy and stroke-free world: using data to inform global action} (\bibinfo{year}{2023}).

\bibitem{rajiah2023cardiac}
\bibinfo{author}{Rajiah, P.~S.}, \bibinfo{author}{Fran{\c{c}}ois, C.~J.} \& \bibinfo{author}{Leiner, T.}
\newblock \bibinfo{title}{Cardiac mri: state of the art}.
\newblock \emph{\bibinfo{journal}{Radiology}} \textbf{\bibinfo{volume}{307}}, \bibinfo{pages}{e223008} (\bibinfo{year}{2023}).

\bibitem{von2023cardiovascular}
\bibinfo{author}{von Knobelsdorff-Brenkenhoff, F.} \& \bibinfo{author}{Schulz-Menger, J.}
\newblock \bibinfo{title}{Cardiovascular magnetic resonance in the guidelines of the european society of cardiology: a comprehensive summary and update}.
\newblock \emph{\bibinfo{journal}{Journal of Cardiovascular Magnetic Resonance}} \textbf{\bibinfo{volume}{25}}, \bibinfo{pages}{42} (\bibinfo{year}{2023}).

\bibitem{10.1093/eurheartj/ehaf192}
\bibinfo{author}{Schulz-Menger, J.} \emph{et~al.}
\newblock \bibinfo{title}{2025 esc guidelines for the management of myocarditis and pericarditis: Developed by the task force for the management of myocarditis and pericarditis of the european society of cardiology (esc)endorsed by the association for european paediatric and congenital cardiology (aepc) and the european association for cardio-thoracic surgery (eacts)}.
\newblock \emph{\bibinfo{journal}{European Heart Journal}} \bibinfo{pages}{ehaf192} (\bibinfo{year}{2025}).
\newblock \urlprefix\url{https://doi.org/10.1093/eurheartj/ehaf192}.

\bibitem{kramer2020standardized}
\bibinfo{author}{Kramer, C.~M.} \emph{et~al.}
\newblock \bibinfo{title}{Standardized cardiovascular magnetic resonance imaging (cmr) protocols: 2020 update}.
\newblock \emph{\bibinfo{journal}{Journal of Cardiovascular Magnetic Resonance}} \textbf{\bibinfo{volume}{22}}, \bibinfo{pages}{17} (\bibinfo{year}{2020}).

\bibitem{bhuva2019multicenter}
\bibinfo{author}{Bhuva, A.~N.} \emph{et~al.}
\newblock \bibinfo{title}{A multicenter, scan-rescan, human and machine learning cmr study to test generalizability and precision in imaging biomarker analysis}.
\newblock \emph{\bibinfo{journal}{Circulation: Cardiovascular Imaging}} \textbf{\bibinfo{volume}{12}}, \bibinfo{pages}{e009214} (\bibinfo{year}{2019}).

\bibitem{bernard2018deep}
\bibinfo{author}{Bernard, O.} \emph{et~al.}
\newblock \bibinfo{title}{Deep learning techniques for automatic mri cardiac multi-structures segmentation and diagnosis: is the problem solved?}
\newblock \emph{\bibinfo{journal}{IEEE transactions on medical imaging}} \textbf{\bibinfo{volume}{37}}, \bibinfo{pages}{2514--2525} (\bibinfo{year}{2018}).

\bibitem{campello2021multi}
\bibinfo{author}{Campello, V.~M.} \emph{et~al.}
\newblock \bibinfo{title}{Multi-centre, multi-vendor and multi-disease cardiac segmentation: the m\&ms challenge}.
\newblock \emph{\bibinfo{journal}{IEEE Transactions on Medical Imaging}} \textbf{\bibinfo{volume}{40}}, \bibinfo{pages}{3543--3554} (\bibinfo{year}{2021}).

\bibitem{martin2023deep}
\bibinfo{author}{Mart{\'\i}n-Isla, C.} \emph{et~al.}
\newblock \bibinfo{title}{Deep learning segmentation of the right ventricle in cardiac mri: the m\&ms challenge}.
\newblock \emph{\bibinfo{journal}{IEEE Journal of Biomedical and Health Informatics}} \textbf{\bibinfo{volume}{27}}, \bibinfo{pages}{3302--3313} (\bibinfo{year}{2023}).

\bibitem{second-annual-data-science-bowl}
\bibinfo{author}{A, N.} \emph{et~al.}
\newblock \bibinfo{title}{Second annual data science bowl} (\bibinfo{year}{2015}).
\newblock \urlprefix\url{https://kaggle.com/competitions/second-annual-data-science-bowl}.

\bibitem{xue2021landmark}
\bibinfo{author}{Xue, H.} \emph{et~al.}
\newblock \bibinfo{title}{Landmark detection in cardiac mri by using a convolutional neural network}.
\newblock \emph{\bibinfo{journal}{Radiology: Artificial Intelligence}} \textbf{\bibinfo{volume}{3}}, \bibinfo{pages}{e200197} (\bibinfo{year}{2021}).

\bibitem{lalande2020emidec}
\bibinfo{author}{Lalande, A.} \emph{et~al.}
\newblock \bibinfo{title}{Emidec: a database usable for the automatic evaluation of myocardial infarction from delayed-enhancement cardiac mri}.
\newblock \emph{\bibinfo{journal}{Data}} \textbf{\bibinfo{volume}{5}}, \bibinfo{pages}{89} (\bibinfo{year}{2020}).

\bibitem{li2023myops}
\bibinfo{author}{Li, L.} \emph{et~al.}
\newblock \bibinfo{title}{Myops: A benchmark of myocardial pathology segmentation combining three-sequence cardiac magnetic resonance images}.
\newblock \emph{\bibinfo{journal}{Medical Image Analysis}} \textbf{\bibinfo{volume}{87}}, \bibinfo{pages}{102808} (\bibinfo{year}{2023}).

\bibitem{bommasani2021opportunities}
\bibinfo{author}{Bommasani, R.} \emph{et~al.}
\newblock \bibinfo{title}{On the opportunities and risks of foundation models}.
\newblock \emph{\bibinfo{journal}{arXiv preprint arXiv:2108.07258}}  (\bibinfo{year}{2021}).

\bibitem{ma2024segment}
\bibinfo{author}{Ma, J.} \emph{et~al.}
\newblock \bibinfo{title}{Segment anything in medical images}.
\newblock \emph{\bibinfo{journal}{Nature Communications}} \textbf{\bibinfo{volume}{15}}, \bibinfo{pages}{654} (\bibinfo{year}{2024}).

\bibitem{sun2025foundation}
\bibinfo{author}{Sun, Y.}, \bibinfo{author}{Wang, L.}, \bibinfo{author}{Li, G.}, \bibinfo{author}{Lin, W.} \& \bibinfo{author}{Wang, L.}
\newblock \bibinfo{title}{A foundation model for enhancing magnetic resonance images and downstream segmentation, registration and diagnostic tasks}.
\newblock \emph{\bibinfo{journal}{Nature Biomedical Engineering}} \textbf{\bibinfo{volume}{9}}, \bibinfo{pages}{521--538} (\bibinfo{year}{2025}).

\bibitem{shad2023generalizable}
\bibinfo{author}{Shad, R.} \emph{et~al.}
\newblock \bibinfo{title}{A generalizable deep learning system for cardiac mri}.
\newblock \emph{\bibinfo{journal}{arXiv preprint arXiv:2312.00357}}  (\bibinfo{year}{2023}).

\bibitem{jacob2024towards}
\bibinfo{author}{Jacob, A.~J.} \emph{et~al.}
\newblock \bibinfo{title}{Towards a vision foundation model for comprehensive assessment of cardiac mri}.
\newblock \emph{\bibinfo{journal}{arXiv preprint arXiv:2410.01665}}  (\bibinfo{year}{2024}).

\bibitem{zhang2025towards}
\bibinfo{author}{Zhang, Y.} \emph{et~al.}
\newblock \bibinfo{title}{Towards cardiac mri foundation models: Comprehensive visual-tabular representations for whole-heart assessment and beyond}.
\newblock \emph{\bibinfo{journal}{arXiv preprint arXiv:2504.13037}}  (\bibinfo{year}{2025}).

\bibitem{caron2021emerging}
\bibinfo{author}{Caron, M.} \emph{et~al.}
\newblock \bibinfo{title}{Emerging properties in self-supervised vision transformers} (\bibinfo{year}{2021}).

\bibitem{he2022masked}
\bibinfo{author}{He, K.} \emph{et~al.}
\newblock \bibinfo{title}{Masked autoencoders are scalable vision learners}.
\newblock \emph{\bibinfo{journal}{Proceedings of the IEEE/CVF conference on computer vision and pattern recognition}}  (\bibinfo{year}{2022}).

\bibitem{zhou2023foundation}
\bibinfo{author}{Zhou, Y.} \emph{et~al.}
\newblock \bibinfo{title}{A foundation model for generalizable disease detection from retinal images}.
\newblock \emph{\bibinfo{journal}{Nature}} \textbf{\bibinfo{volume}{622}}, \bibinfo{pages}{156--163} (\bibinfo{year}{2023}).

\bibitem{bachmann2022multimae}
\bibinfo{author}{Bachmann, R.}, \bibinfo{author}{Mizrahi, D.}, \bibinfo{author}{Atanov, A.} \& \bibinfo{author}{Zamir, A.}
\newblock \bibinfo{title}{Multimae: Multi-modal multi-task masked autoencoders} (\bibinfo{year}{2022}).

\bibitem{gao2022convmae}
\bibinfo{author}{Gao, P.} \emph{et~al.}
\newblock \bibinfo{title}{Convmae: Masked convolution meets masked autoencoders}.
\newblock \emph{\bibinfo{journal}{arXiv preprint arXiv:2205.03892}}  (\bibinfo{year}{2022}).

\bibitem{raisi2021cardiovascular}
\bibinfo{author}{Raisi-Estabragh, Z.}, \bibinfo{author}{Harvey, N.~C.}, \bibinfo{author}{Neubauer, S.} \& \bibinfo{author}{Petersen, S.~E.}
\newblock \bibinfo{title}{Cardiovascular magnetic resonance imaging in the uk biobank: a major international health research resource}.
\newblock \emph{\bibinfo{journal}{European Heart Journal-Cardiovascular Imaging}} \textbf{\bibinfo{volume}{22}}, \bibinfo{pages}{251--258} (\bibinfo{year}{2021}).

\bibitem{hatamizadeh2022unetr}
\bibinfo{author}{Hatamizadeh, A.} \emph{et~al.}
\newblock \bibinfo{title}{Unetr: Transformers for 3d medical image segmentation} (\bibinfo{year}{2022}).

\bibitem{targ2016resnet}
\bibinfo{author}{Targ, S.}, \bibinfo{author}{Almeida, D.} \& \bibinfo{author}{Lyman, K.}
\newblock \bibinfo{title}{Resnet in resnet: Generalizing residual architectures}.
\newblock \emph{\bibinfo{journal}{arXiv preprint arXiv:1603.08029}}  (\bibinfo{year}{2016}).

\bibitem{seyyed2020chexclusion}
\bibinfo{author}{Seyyed-Kalantari, L.}, \bibinfo{author}{Liu, G.}, \bibinfo{author}{McDermott, M.}, \bibinfo{author}{Chen, I.~Y.} \& \bibinfo{author}{Ghassemi, M.}
\newblock \bibinfo{title}{Chexclusion: Fairness gaps in deep chest x-ray classifiers}.
\newblock \emph{\bibinfo{journal}{BIOCOMPUTING 2021: proceedings of the Pacific symposium}} \bibinfo{pages}{232--243} (\bibinfo{year}{2020}).

\bibitem{dai2024deepseekmoe}
\bibinfo{author}{Dai, D.} \emph{et~al.}
\newblock \bibinfo{title}{Deepseekmoe: Towards ultimate expert specialization in mixture-of-experts language models}.
\newblock \emph{\bibinfo{journal}{arXiv preprint arXiv:2401.06066}}  (\bibinfo{year}{2024}).

\bibitem{isensee2021nnu}
\bibinfo{author}{Isensee, F.}, \bibinfo{author}{Jaeger, P.~F.}, \bibinfo{author}{Kohl, S.~A.}, \bibinfo{author}{Petersen, J.} \& \bibinfo{author}{Maier-Hein, K.~H.}
\newblock \bibinfo{title}{nnu-net: a self-configuring method for deep learning-based biomedical image segmentation}.
\newblock \emph{\bibinfo{journal}{Nature methods}} \textbf{\bibinfo{volume}{18}}, \bibinfo{pages}{203--211} (\bibinfo{year}{2021}).

\bibitem{mcintosh2024isambard}
\bibinfo{author}{McIntosh-Smith, S.}, \bibinfo{author}{Alam, S.~R.} \& \bibinfo{author}{Woods, C.}
\newblock \bibinfo{title}{Isambard-ai: a leadership class supercomputer optimised specifically for artificial intelligence}.
\newblock \emph{\bibinfo{journal}{arXiv preprint arXiv:2410.11199}}  (\bibinfo{year}{2024}).

\bibitem{feng2018wing}
\bibinfo{author}{Feng, Z.-H.}, \bibinfo{author}{Kittler, J.}, \bibinfo{author}{Awais, M.}, \bibinfo{author}{Huber, P.} \& \bibinfo{author}{Wu, X.-J.}
\newblock \bibinfo{title}{Wing loss for robust facial landmark localisation with convolutional neural networks} (\bibinfo{year}{2018}).

\end{thebibliography}

\end{document}